\documentstyle[prd,aps,epsfig,twocolumn]{revtex}
\begin{document}
\draft

\title{Photon--meson transition form factors $\gamma\pi^0$,
$\gamma\eta$ and $\gamma\eta'$ at low and moderately high $Q^2$}
\author{V.V.Anisovich\thanks{Electronic address: anisovic@lnpi.spb.su}
, D.I.Melikhov\thanks{Also at {\it Nuclear Physics Institute, Moscow
State University}, Electronic address: melikhov@monet.npi.msu.su},
and V.A.Nikonov\thanks{Electronic address: nikon@rec03.pnpi.spb.ru}}
\address{St.Petersburg Nuclear Physics Institute, Gatchina, 188350,
Russia}
\date{\today}
\maketitle
  \widetext
\begin{abstract}
We study photon--meson transition form factors
$\gamma^*(Q^2)\gamma\to\pi^0,\eta,\eta'$ at low and moderately high
virtualities $Q^2$ of one of the photons, the second photon being
real. For the description of the form factor at low $Q^2$, a
nontrivial quark--antiquark $(q\bar q)$ structure of the photon in the
soft region is assumed, i.e. the photon is treated much like an
ordinary vector meson. At large $Q^2$, along with a perturbative tail
of the soft wave function, as it is for a hadron, the photon wave
function contains also a standard QED point-like $q\bar q$ component.
The latter provides the $1/Q^2$ behavior of the transition form factor
at large $Q^2$ in accordance with perturbative QCD. Using the
experimental results on $\gamma\pi^0$ form factor, we reconstruct the
soft photon wave function which is found to have the structure similar
to pion's, which has been formerly determined from a study of the
elastic pion form factor.  Assuming the universality of the
ground--state pseudoscalar meson wave functions we calculate the
transition form factors $\gamma\eta$, $\gamma\eta'$ and partial widths
$\eta\to\gamma\gamma$, $\eta'\to\gamma\gamma$, in a perfect agreement
with data.
\end{abstract}


\narrowtext
\section{Introduction} \label{sec-intro}

In the present paper we continue the study of the transition regime
from soft nonperturbative physics to the physics of hard processes
described by the perturbative QCD (PQCD). In Ref. \cite{amn} we have
proposed a method to consider a form factor in a broad range of
momentum transfers starting from the nonperturbative region of small
$Q^2$ and moving to moderately large $Q^2$ where form factor is
represented as a series in $\alpha_s$ accounting for the
nonperturbative  term and $O(\alpha_s)$ corrections. This allows a
continuous transition from small to asymptotically large momentum
transfers. Within this procedure, we have determined the pion
light--cone wave function by describing the pion elastic form factor
in the range $0\le Q^2\le 10\;GeV^2$.

Recent experiments on pseudoscalar meson production in $e^+e^-$
collisions \cite{exp} provide new data on the transition form factors
$\gamma^*(Q^2)\gamma\to \pi^0,\eta,\eta'$ in the region of the
momentum transfers $0\le Q^2\le 20\;GeV^2$. These results open a new
possibility for studying the onset of the asymptotical PQCD regime.
Reconstruction of the pseudoscalar meson wave function and photon
$q\bar q$-distribution  provides a bridge between the phenomenological
soft-physics description and rigorous results of PQCD.

Theoretical investigation of the photon--meson transition processes
which has a long history has given two important results on the
photon--pion transition form factor:

  $$\quad\vspace{4cm}$$

(1) Adler--Bell--Jackiw axial anomaly \cite{abj} yields nonvanishing
transition form factor of the pion into two real photons in the chiral
limit of vanishing quark masses
\begin{equation} \label{fzero}
F_{\gamma^*\gamma^*\pi}(Q_1^2=0,\;Q_2^2=0)=\frac1{2\sqrt{2}\pi^2
f_\pi}\ , \; f_\pi=130\;MeV\ ,
\end{equation}
where the photon--pion transition form factor is defined as follows
($Q_1^2=-q_1^2\ ,\; Q_2^2=-q_2^2$):
\begin{eqnarray}\label{fdef}
\langle\pi(P)|&T&|\gamma(q_1,\mu)\gamma(q_2,\nu)\rangle\; \nonumber\\
=&&e^2\varepsilon_{\mu\nu\alpha_1\alpha_2}q_1^{\alpha_1}q_2^{\alpha_2}
F_{\gamma^*\gamma^*\pi}(-q_1^2, -q_2^2).
\end{eqnarray}

(2) In the kinematical region where at least one of the photon
virtualities is large, PQCD gives the following prediction for the
behavior of the transition form factor \cite{bl}
\begin{equation}\label{fass}
F_{\gamma^*\gamma^*\pi}(Q_1^2\ ,\; Q_2^2)=\frac{\sqrt2}{3}
\int\limits_0^1\frac{\phi_\pi(x)dx}{xQ_1^2+(1-x)Q_2^2}\ ,
\end{equation}
where $\phi_\pi(x)$ is the leading twist wave function (distribution
amplitude) which describes the longitudinal momentum distribution of
valence quark--antiquark pair in the pion. PQCD also predicts
asymptotic behavior of the pion distribution amplitude in the
following form \cite{hsp}:
\begin{equation}
\label{phias}
\phi^{as}_\pi(x)=6f_\pi x(1-x).
\end{equation}
For  $Q_2^2\to 0$ and large $Q_1^2\equiv Q^2$ which correspond to
realistic kinematics of the experiments \cite{exp}, Eq.\ (\ref{fass})
gives
\begin{equation}
F_{\gamma\pi}(Q^2)
=\frac{\sqrt2}3\int\limits_0^1\frac{\phi_\pi(x) dx}{xQ^2}
\left[1+O(\alpha_s(Q^2)\right] +O\left(\frac 1{Q^4}\right),
\end{equation}

\newpage
\widetext
\begin{figure}
\centerline{\epsfig{file=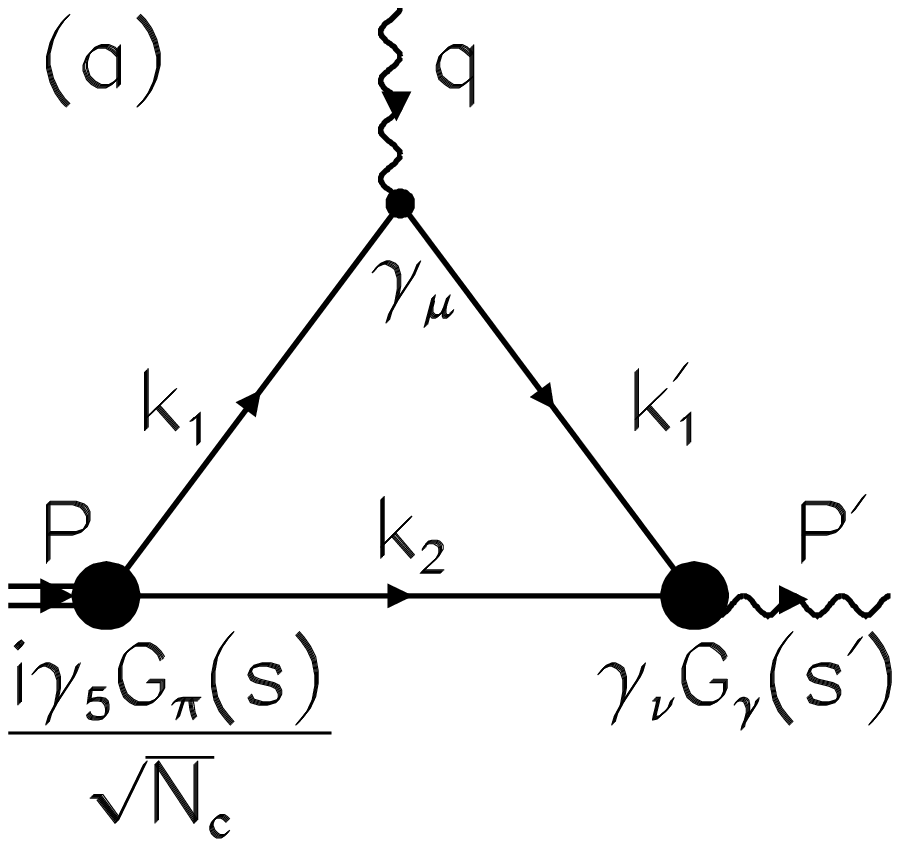,width=4.4cm}
            \epsfig{file=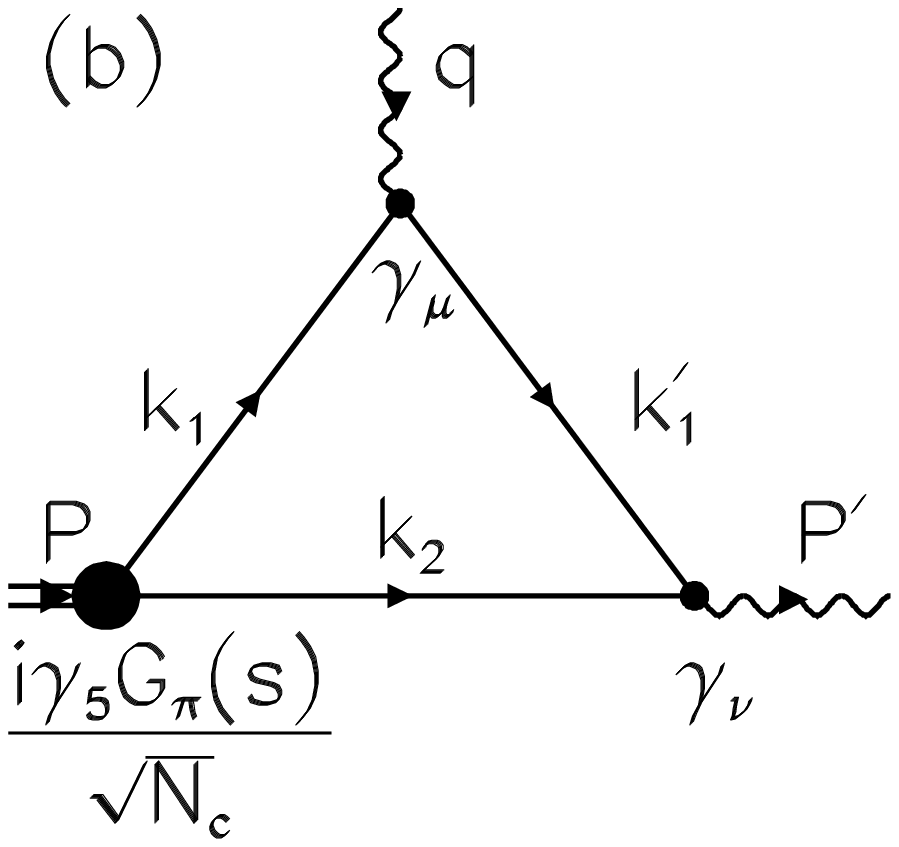,width=4.4cm}
            \epsfig{file=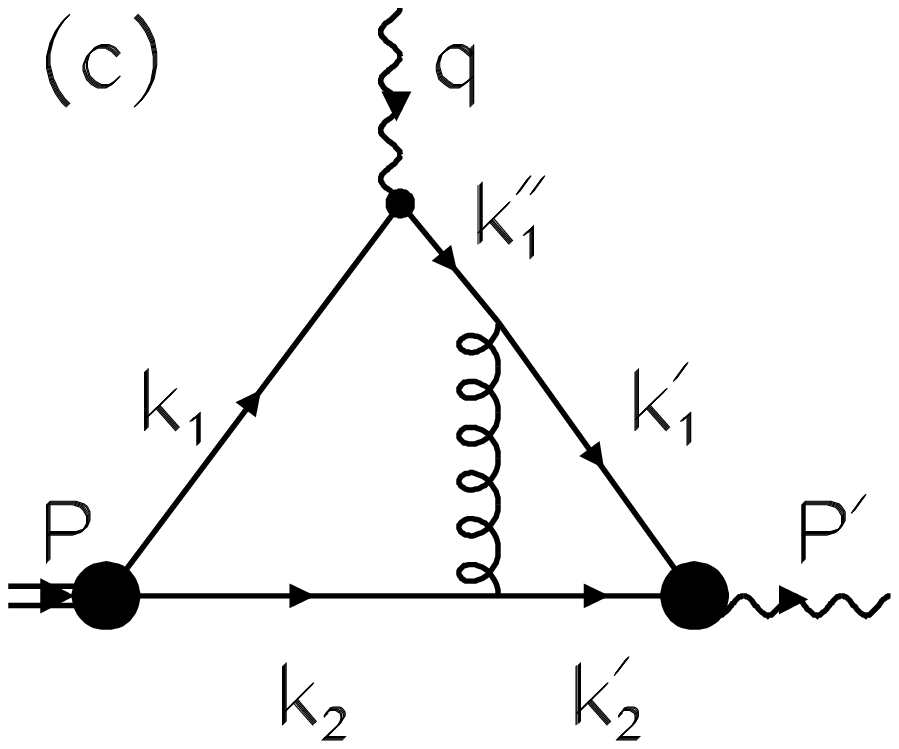,width=4.4cm}
            \epsfig{file=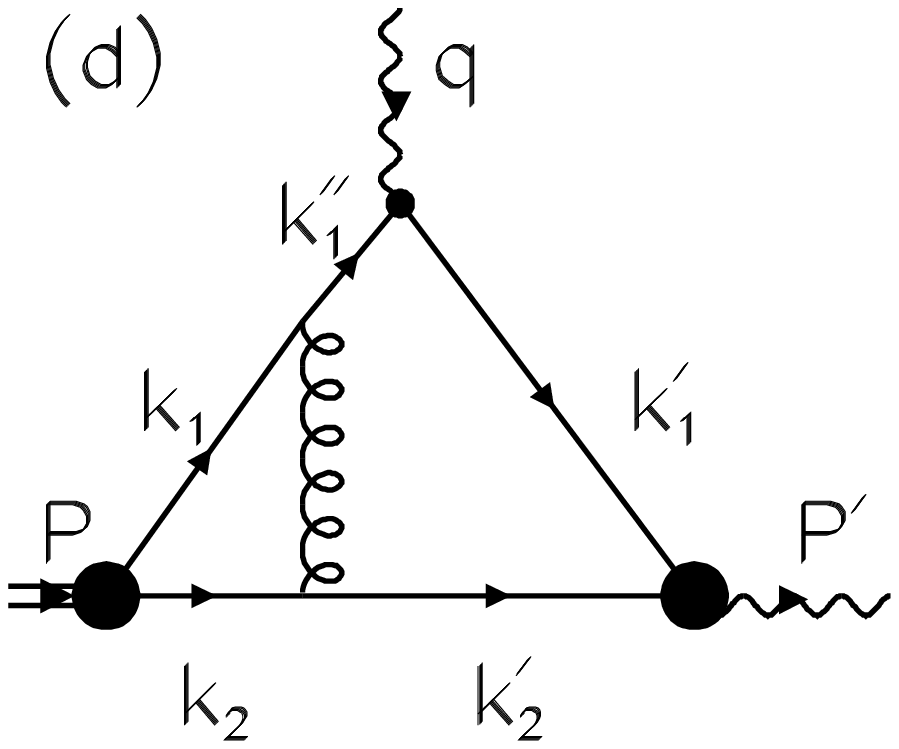,width=4.4cm}}
\caption{Diagrams relevant to the description of the transition
form factor at low and moderately high $Q^2$.}
\label{fig:series}
\end{figure}
\narrowtext
\noindent
where $F_{\gamma\pi}(Q^2)\equiv F_{\gamma^*\gamma^*\pi}(Q^2,0)$.

At asymptotically large $Q^2$ one can use the asymptotic pion
distribution amplitude to find the leading behavior of the transition
form factor
\begin{equation}
F_{\gamma\pi}(Q^2)=
\frac{\sqrt{2}f_\pi}{Q^2}\left[1+O(\alpha_s(Q^2)\right]+O\left(\frac
1{Q^4}\right).
\end{equation}
The leading term corresponds to the diagram of Fig.\ \ref{fig:series}b
with point-like vertices of the quark--photon interaction.

A problem starting from which $Q^2$ the form factor can be reliably
described by the leading PQCD term only has been extensively discussed
in connection with  pion elastic form factor \cite{ils}, \cite{all}.
As was firstly underlined in Ref. \cite{ils}, asymptotical regime is
switched on at considerably large momentum transfers, whereas in the
region of $Q^2\approx 10-20\;GeV^2$ the soft nonperturbative wave
function gives a substantial contribution.  Quantitative results for
the pion form factor \cite{amn} agree with the statements of Ref.
\cite{ils}.

The procedure of Ref. \cite{amn} is the following: we divide the pion
light--cone wave function $\Psi^\pi$ into soft and hard components,
$\Psi_S^\pi$ and $\Psi_H^\pi$, such as $\Psi_S^\pi$ is large at
$s=(m^2+\vec k_\perp^2)/(x(1-x))<s_0$, while $\Psi_H^\pi$ prevails at
$s>s_0$. We performed this decomposition of the wave function using
the step--function as a simplest ansatz
\begin{equation}
\label{psidec}
\Psi^\pi = \Psi_S^\pi\theta(s_0-s) + \Psi_H^\pi \theta(s-s_0).
\end{equation}
It is reasonable to represent the hard component, $\Psi_H^\pi$, as a
series in $\alpha_s$. Then at small and intermediate $Q^2$ the elastic
pion form factor reads
\begin{equation} \label{ffdec}
F_\pi=F^{SS}_\pi+2F^{SH}_\pi+O(\alpha^2_s),
\end{equation}
where $F^{SS}_\pi$ is a truly nonperturbative part of the form factor,
and $F^{SH}_\pi$ is an $O(\alpha_s)$ term with one--gluon exchange.
The first term clearly dominates in the pion form factor at small
$Q^2$. With minor corrections at small $Q^2$, the second term, as is
known, provides the leading $\alpha_s(Q^2)/Q^2$ behavior of the
elastic form factor at asymptotically large $Q^2$. So, the first two
terms in the right-hand side of Eq.\ (\ref{ffdec}) accumulate the
leading behavior from the two regions of small and large $Q^2$, thus
allowing a realistic description in the region of intermediate
momentum transfers.

  $$\quad\vspace{3.8cm}$$

Hard wave function, $\Psi_H^\pi$, is represented as a convolution of
the one-gluon exchange kernel $V^{\alpha_s}$ with $\Psi_S$
\begin{equation}
\label{psihard}
\Psi^\pi_H=V^{\alpha_s}\otimes \Psi^\pi_S.
\end{equation}

So, soft pion wave function $\Psi^\pi_S$ is responsible for pion form
factor behavior both at small and moderately large $Q^2$.

This strategy, with the incorporation of "photon wave function", is
applicable for the description of the photon--pion transition form
factor:
\begin{equation} \label{psigam}
\Psi^\gamma=\Psi_S^\gamma\theta(s_0-s)+\Psi_H^\gamma \theta(s-s_0).
\end{equation}

However, an important distinction of the photon hard wave function
compared with that of a  hadron should be taken into account: namely,
in addition to the  hadronic component of the hard wave function which
is related to the soft wave function via Eq.\ (\ref{psihard}), the
photon hard wave function contains also a standard point-like QED
$q\bar q$--component such as
\begin{equation}
\label{psihardg}
\Psi^\gamma_H=V^{\alpha_s}\otimes\Psi^\gamma_S+\Psi^\gamma_{pt}.
\end{equation}

Corresponding expression for the form factor takes the form (Fig.
\ref{fig:series}):
$$
F_{\gamma\pi}=\Psi^\pi_S\bullet \Psi^\gamma_S+\Psi^\pi_S\bullet
\Psi^\gamma_H+\Psi^\pi_H\bullet \Psi^\gamma_S
$$
$$
=\Psi^\pi_S\bullet \Psi^\gamma_S +\Psi^\pi_S\bullet \Psi^\gamma_{pt}
+\Psi^\pi_S\bullet V^{\alpha_s}\otimes\Psi^\gamma_S
+\Psi^\pi_S\otimes V^{\alpha_s}\bullet \Psi^\gamma_S
$$
\FL \begin{equation}\label{ffdecg}
\equiv F^{SS}+F^{Spt}+F^{SH(1)}+F^{SH(2)}\ .
\end{equation}

Our analysis shows that at small momentum transfers the $F^{SS}$ part
dominates in the transition form factor, i.e. in the soft region
photon should be treated much like an ordinary hadron, just in the
spirit of the vector meson dominance. At large $Q^2$, the soft--point
term ($F^{Spt}$), gives the leading $1/Q^2$ falloff, whereas the
contribution of the $O(\alpha_s)$ terms ($F^{SH(1)}+F^{SH(2)}$) is
suppressed by the additional factor $\alpha_s$: the behavior of the
photon--pion transition form factor differs from that of the elastic
pion form factor where the soft--point term is absent and the
soft--hard terms are dominant.

At intermediate momentum transfers, Eq.\ (\ref{ffdecg}) provides
substantial corrections to the $1/Q^2$ falloff: namely, these
corrections are due to the transverse motion of quarks in the
soft--point term as well as to contributions related to the nontrivial
$q\bar q$--structure of the soft photon (the terms involving
$\Psi^\gamma_S$). The allowance for the corrections of the first type
is a standard procedure in moving from asymptotically large to
intermediate momentum transfers within the hard scattering approach
\cite{huang} and modified hard scattering approach \cite{krollong}
which are based on the account for transverse motion and Sudakov
effects \cite{ls}. However, as the present analysis shows, in the
region of a few GeV$^2$ the contribution of the terms originated from
the nontrivial $q\bar q$--structure of the soft photon is substantial
and cannot be neglected.

The proposed method allows a self--consistent description of the
photon--pion transition form factor in a broad range of $Q^2$. We use
the soft pion wave function which has been previously determined in
Ref. \cite{amn} by fitting  elastic pion form factor. The description
of the photon--pion transition form factor reveals a similarity of the
low--$s$ $q\bar q$-structure of soft photon with that of a pion.

Summing up, we obtain the following results:

1. We fit the available data on the photon--pion transition form
factor at $Q^2=0-8\;GeV^2$ and determine the soft $q\bar q$ wave
function of the photon. Pion wave function is taken from our
description of elastic pion form factor \cite{amn}. Soft photon wave
function turns out to be close to the ground--state meson wave
function at low $s$, i.e. $\Psi_S^{\gamma}(s)\sim\Psi_S^{\pi}(s)$ at
$s\leq 2\; GeV^2$. The similarity of $\Psi_S^{\gamma}$ and
$\Psi_S^{\pi}$ at $s\leq 2\; GeV^2$ seems to be quite natural and
corresponds to the vector meson dominance in the vertex $\gamma\to
q\bar q$. At $s=s_0$ the photon wave function satisfies the boundary
condition $\Psi_S^\gamma(s_0)=\Psi_{pt}^\gamma(s_0)$, which provides a
correct sewing of the $F^{SS}$ and $F^{Spt}$ terms. Soft wave
functions determined, we calculate the photon--pion transition form
factor in a broad range of $Q^2$.  Calculations show that several
kinematical regions, where different contributions to the transition
form factor dominate, may be isolated:

(i) At small $Q^2=0-5\;GeV^2$ the transition form factor is dominated
by the soft--soft term which corresponds to nontrivial hadron--like
structure of the soft photon.

(ii) At large $Q^2\ge 50\;GeV^2$ the QED point-like component of the
photon gives the main contribution reproducing the PQCD result.

(iii) In the intermediate region the transition form factor is an
interplay of the soft--soft, soft--point and soft--hard contributions.
Numerically, the form factor behavior is very close to the
interpolation formula proposed by Brodsky and Lepage \cite{bl}:
\begin{equation}
F_{\gamma\pi}(Q^2)=\frac{\sqrt2 f_\pi}{Q^2+4\pi^2f_\pi^2}.
\end{equation}

2. We calculate $\gamma\eta$ and $\gamma\eta'$ transition form
factors, assuming universality of the $\vec k$-dependence of  wave
functions of all the ground--state pseudoscalar mesons where $\vec k$
is relative momentum of constituent quarks, $s=4m^2+4\vec k^2$. This
assumption is in line with the conventional quark model. In accordance
with vector meson dominance, the same ansatz is used for relating the
nonstrange and strange components of the soft photon. Then
$\gamma\eta$ and $\gamma\eta'$ form factors are calculated with no
free parameters. The results for the $\gamma\eta$ and $\gamma\eta'$
transition form factors are in an excellent agreement with the
experimental data both on the shape of the form factors and on the
decay partial widths, $\Gamma(\eta\to\gamma\gamma)$ and
$\Gamma(\eta'\to\gamma\gamma)$. It provides an argument for small
admixture of a glueball (two gluon) component into $\eta$ and $\eta'$:
within the experimental accuracy we estimate corresponding
probabilities as $W_\eta(glueball)<10\%$ and
$W_{\eta'}(glueball)<20\%$.

The paper is organized as follows: \\ In Section \ref{sec-piff} the
calculation of the $\gamma\pi^0$ transition form factor is performed
and the soft transition vertex $\gamma\to q\bar q$ is reconstructed.
Section \ref{sec-etaff} is devoted to the calculation of $\gamma\eta$
and $\gamma\eta'$ transition form factors. Conclusive remarks are
given in Section \ref{sec-con}. Appendix presents the calculation
details.

\section{$\gamma\pi^0$ transition form factor} \label{sec-piff}

We consider the $\gamma\pi^0$ transition form factor using the method
presented in detail in Ref. \cite{amn}. So we omit here a discussion
of the basic points of the technique, outlining only a skeleton of the
calculation procedure.

The form factor $F_{\gamma\pi}$ is connected with the amplitude of the
process $\gamma^*\gamma\to\pi^0$ as follows:
\begin{equation}
T^{\gamma^*\gamma\pi}_{\mu\nu}(q^2)=e^2
\varepsilon_{\mu\mu\alpha\beta}q^\alpha P^\beta F_{\gamma\pi}(-q^2),
\end{equation}
where $P$ is the pion momentum. Then partial width of the decay
$\pi^0\to\gamma\gamma$ reads
\begin{equation}\label{decrate}
\Gamma(\pi^0\to\gamma\gamma)=
\frac\pi4\alpha^2m_{\pi}^3F^2_{\gamma\pi}(0),\quad \alpha=1/137.
\end{equation}

\subsection{Soft--soft term $F_{\gamma\pi}^{SS}(Q^2)$}
\label{sec-ss}

$F^{SS}_{\gamma\pi}(Q^2)$ corresponds to the diagram of Fig.
\ref{fig:series}a. The vertices of the soft transitions $\pi\to q\bar
q$ and $q\bar q\to\gamma$ are defined as
\begin{equation}
\frac{\bar q i\gamma_5 q}{\sqrt{N_c}} G_\pi (s)
\end{equation}
for $pion\to quark+antiquark$ ($N_c$ is the number of colors) and
\begin{equation}
e_q\bar q \gamma_\nu q G_\gamma (s')\
\end{equation}
for $q\bar q\to photon$ ($e_q$ is the quark charge).

The contribution of the diagram Fig.\ \ref{fig:series}a can be
written as the following double dispersion representation:
\begin{eqnarray}\label{dispss}
F_{\gamma\pi}^{SS}(Q^2)&=&Z_\pi 2 f_q(Q^2)\sqrt{N_c}
\int\frac{ds G_\pi(s)\theta(s_0-s)}{\pi (s-m_\pi^2)}\nonumber \\
&\times&
\frac{ds' G_\gamma(s')\theta(s_0-s')}{\pi s'}
\Delta_{\pi\gamma^*\gamma}(s,s',Q^2).
\end{eqnarray}
Here $f_q(Q^2)$ is quark form factor, $Z_\pi$ is a charge factor:
$Z_\pi=(e_u^2-e_d^2)/\sqrt 2$, with $e_u=2/3$ and $e_d=-1/3$.

The quantity $\Delta_{\pi\gamma^*\gamma}(s,s',Q^2)$ is the double
spectral density of Feynman diagram of Fig.\ \ref{fig:series}a
with point-like vertices and the off--shell pion and photon momenta:
\begin{eqnarray} \label{Ftriangle}
&&
\varepsilon_{\mu\mu\alpha\beta}q^\alpha P^\beta
\Delta_{\pi\gamma^*\gamma}(s,s',Q^2)=-disc_s disc_{s'}
\nonumber \\
&&
\int\frac{d^4k_2}{i(2\pi)^4}
\frac{Sp\left({i\gamma_5(m-\hat k_2)\gamma_\nu(m+\hat k_1')
\gamma_\mu(m+\hat k_1)}\right)}
{(m^2-k_1^2)(m^2-k_2^2)(m^2-k_1'^2)}, \nonumber \\
&&
s=(k_1+k_2)^2\ ,\quad s'=(k_1'+k_2)^2\ .
\end{eqnarray}
$m$ is the non--strange quark mass.

The trace reads:
\begin{eqnarray} \label{Spur1}
Sp\left(i\gamma_5(m-\hat k_2)\gamma_\nu(m+\hat k_1')\right.
&\gamma_\mu&\left.(m+\hat k_1)\right)= \nonumber \\
&-&4m\varepsilon_{\mu\mu\alpha\beta}q^\alpha P^\beta.
\end{eqnarray}
Then one finds
\begin{eqnarray}
&&
\varepsilon_{\mu\mu\alpha\beta}q^\alpha P^\beta
\Delta_{\pi\gamma^*\gamma}(s,s',Q^2)=\frac {(-i)^2(2\pi i)^3}{4}
4m\varepsilon_{\mu\mu\alpha\beta}q^\alpha P^\beta\nonumber \\
&&
\times
\int\frac{d^4k_2}{i(2\pi)^4} \delta(m^2-k_1^2)\delta(m^2-k_2^2)
\delta(m^2-k_1'^2)\ , \nonumber \\
&&
s=(k_1+k_2)^2\ ,\quad s'=(k_1'+k_2)^2\ ,
\end{eqnarray}
where
\begin{equation}\label{delta}
\Delta_{\pi\gamma^*\gamma}(s,s',Q^2)=\frac m4
\frac{\theta(s'sQ^2-m^2\lambda(s,s',Q^2))}{\lambda^{1/2}(s,s',Q^2)},
\end{equation}
and
\begin{equation}\label{lambda}
\lambda(s,s',Q^2)=(s'-s)^2+2Q^2(s'+s)+Q^4. \nonumber
\end{equation}

Eqs.\ (\ref{dispss}), (\ref{delta}) and (\ref{lambda}) present
$F_{\gamma\pi}^{SS}(Q^2)$ in terms of invariant variables $s$ and
$s'$. In the soft--soft term these variables are constrained within
the intervals $4m^2\le s \le s_0$ and $4m^2\le s' \le s_0$.

Introducing the light--cone variables, $x=k_{2+}/P_+$, $\vec k_\perp
=\vec k_{2\perp}$ and $q=(0,q_-,\vec Q )$ (see Eq.\ (\ref{lccom})),
one finds
\begin{eqnarray}\label{deltalc}
&&
\Delta_{\pi\gamma^*\gamma}(s,s',Q^2)=\frac m{4\pi}
\int \frac{dxd^2\vec k_\perp }{x(1-x)^2} \nonumber \\
&&
\times
\delta\left(s -\frac{m^2+\vec k_\perp^2 }{x(1-x )}\right)
\delta\left(s'-\frac{m^2+(\vec k_\perp -x\vec Q )^2}{x(1-x)}\right)
\end{eqnarray}
Using these variables one comes to the following expression for the
soft--soft form factor
\begin{eqnarray}\label{finsslc}
F_{\gamma\pi}^{SS}(Q^2)&=&
2Z_\pi f_q(Q^2)\sqrt{N_c}\frac m{4\pi^5}
\int \frac{dxd^2\vec k_\perp }{x(1-x)^2} \nonumber \\
&\times&\Psi_\pi(s)\theta(s_0-s)\frac{G_\gamma(s')}{s'}\theta(s_0-s'),
\end{eqnarray}
where
\begin{equation}\label{psipi}
\Psi_\pi(s)=\frac{G_\pi(s)}{s-m_\pi^2}
\end{equation}
and
\begin{equation}\label{slc}
s=\frac{m^2+\vec k_\perp^2 }{x(1-x )}, \;
s'=\frac{m^2+(\vec k_\perp -x\vec Q )^2}{x(1-x)}.
\end{equation}

The value $F_{\gamma\pi}^{SS}(0)$ is connected with the known
$\pi^0\to\gamma\gamma$ decay width through Eq.\ (\ref{decrate}). On
the other hand, one finds

\begin{equation} \label{gnorm}
F_{\gamma\pi}^{SS}(0)=Z_\pi\frac m{2\pi}\int\limits_{4m^2}^{s_0}
\frac{ds}{\pi}\Psi_\pi(s)\frac{G_\gamma(s)}s
ln\frac{1+\sqrt{1-4m^2/s}}{1-\sqrt{1-4m^2/s}}.
\end{equation}
We use this equation as a normalization condition for $G_{\gamma}(s)$.

$F_{\gamma\pi}^{SS}$ involves the constituent quark form factor which
satisfies the condition $f_q(0)=1$ and turns into  Sudakov's form
factor at large $Q^2$. The Sudakov form factor is taken in the form
\begin{equation}
S(Q^2)=exp\left({-\frac{\alpha_s(Q^2)}{2\pi}C_F
ln^2 \left( {\frac{Q^2}{Q^2_0}} \right) }\right),\;
C_F=\frac{N_c^2-1}{2N_c},
\end{equation}
where $Q_0$ is of the order of $1\;GeV$; we put $Q_0=1\;GeV$.
Coupling constant $\alpha_s(Q^2)$ is assumed to be frozen below
$1\;GeV^2$, namely we set
\begin{equation}\label{alpha}
\alpha_s(Q^2)=\left\{\begin{array}{cr}
const\ , &Q<1\;GeV\ ,\\
 \frac{4\pi}9 ln^{-1}\left(\frac{Q^2}{\Lambda^2}\right)\ ,
&Q>1\;GeV\ ,\end{array}\right.
\end{equation}
where $\Lambda=220\; MeV$.

So, the constituent quark form factor is taken as
\begin{equation}\label{fq}
f_q(Q^2)=\left\{\begin{array}{cr}
      1\ , &Q<Q_0\ , \\
 S(Q^2)\ , &Q>Q_0\ .\end{array}\right.
\end{equation}

Let us stress that we have used here the same quark form factor as in
Ref. \cite{amn}.

\subsection{Soft--point term $F_{\gamma\pi}^{Spt}(Q^2)$}
\label{sec-spt}

Contribution from the diagram of Fig.\ \ref{fig:series}b with
$s'\ge s_0$ is denoted as $F_{\gamma\pi}^{Spt}(Q^2)$ and equals to:
\begin{eqnarray} \label{finsp}
F_{\gamma\pi}^{Spt}(Q^2)&=&
2Z_\pi f_q(Q^2)\sqrt{N_c}\frac m4 \nonumber \\
&\times&
\int\frac{ds}\pi\frac{ds'}{\pi s'} \Psi_\pi(s)
\frac{\theta(s'sQ^2-m^2\lambda(s,s',Q^2))}
{\lambda^{1/2}(s,s',Q^2)}, \nonumber \\
4m^2\le &s& \le s_0,\quad s' \ge s_0.
\end{eqnarray}

In the used normalization of the vertex, the point-like interaction is
defined as $e_q\bar q\gamma_\nu q$, so the sewing of the soft--soft
and soft--point terms requires
\begin{equation}
\label{sewing}  G_\gamma(s_0)=1\ .
\end{equation}
Notice that $F_{\gamma\pi}^{Spt}(0)=0$ which is the consequence of
the constraints $s\le s_0$ and $s'\ge s_0$.

In terms of light--cone variables the expression for the
$F_{\gamma\pi}^{Spt}(Q^2)$ reads
\begin{eqnarray}\label{finsplc}
&&F_{\gamma\pi}^{Spt}(Q^2)=
2Z_\pi f_q(Q^2)\sqrt{N_c}\frac m{4\pi^5}
\int \frac{dxd^2\vec k_\perp }{x(1-x)^2}\Psi_\pi(s)\frac 1{s'}
\nonumber \\
&&\times\theta\left(s_0-\frac{m^2+\vec k_\perp^2 }{x(1-x )}\right)
\theta\left(\frac{m^2+(\vec k_\perp -x\vec Q )^2}{x(1-x)}-s_0\right),
\end{eqnarray}
with $s$ and $s'$ given by Eq.\ (\ref{slc}).

\subsection{Soft--hard term $F_{\gamma\pi}^{SH(1)}(Q^2)$}
\label{sec-sh1}

Soft--hard form factor $F_{\gamma\pi}^{SH(1)}(Q^2)$ is connected
with the amplitude of the diagram of Fig.\ \ref{fig:series}c which can
be written as the following dispersion representation
\begin{eqnarray}
&T&_{\mu\nu}^{SH(1)}(Q^2)=2e^2 Z_\pi \sqrt{N_c} \int
\frac{ds ds' ds''}{\pi\;\pi\;\pi}
\Psi_\pi(s)\frac{G_\gamma(s')}{s'}\frac 1{s''}
\nonumber \\
&\times&D_{\mu\nu}^{(1)}(s,s',s'',Q^2)\theta(s_0-s)
\theta(s_0-s')\theta(s''-s_0).
\end{eqnarray}

In this expression $D_{\mu\nu}^{(1)}(s,s',s'',Q^2)$ is the spectral
density of the diagram of Fig.\ \ref{fig:series}c:
\begin{eqnarray}
D_{\mu\nu}^{(1)}(s,s',s'',Q^2)&&=C_F (2\pi i)^5 \frac{(-i)^3}8
\int \frac{d^4k_2}{i(2\pi)^4} \frac{d^4k'_2}{i(2\pi)^4} \nonumber \\
\times\frac{4\pi\alpha_s(t)}{m_G^2-t} Sp_{\mu\nu}^{(1)}&&
\delta(m^2-k_1^2) \delta(m^2-k_2^2) \delta(m^2-k_1'^2) \nonumber \\
\times&&\delta(m^2-k_2'^2) \delta(m^2-k_1''^2)\ , \nonumber\\
s=(k_1+k_2)^2,\; s'&&=(k_1'+k_2')^2,\; s''=(k_1''+k_2)^2.
\end{eqnarray}
Here $t=(k'_2-k_2)^2$ and $m_G$ is an effective gluon mass; we
consider several choices for $m_G$.

$Sp_{\mu\nu}^{(1)}$ denotes the following trace:
\begin{eqnarray}
&&Sp_{\mu\nu}^{(1)}=Sp\left({i\gamma_5(m-\hat k_2)
\gamma_\alpha(m-\hat k_2')\gamma_\nu(m+\hat k_1')}
\gamma_\alpha \right. \nonumber\\
&&\hspace{3.8cm}\left.
\times(m+\hat k_1'')\gamma_\mu(m+\hat k_1)\right)\ .
\end{eqnarray}

For a detailed calculation of the quantity $Sp_{\mu\nu}^{(1)}$ we
refer to Appendix. Isolating the factor
$\varepsilon_{\mu\mu\alpha\beta}q^\alpha P^\beta $ one finds
$$
Sp_{\mu\nu}^{(1)}=\varepsilon_{\mu\mu\alpha\beta}q^\alpha P^\beta
\cdot S(1)
$$
\begin{equation}
S(1)=4m\left[s'-12b'+4b''+(s''-s+Q^2)(a_1'-a_2')\right].
\end{equation}
Analytic expressions for $a_1',a_2',b',b''$ are
given in Appendix (see Eqs.\ (\ref{kmu}), (\ref{acoefs}) and
(\ref{bcoefs})).

Thus, one finds
$$
D_{\mu\nu}^{(1)}(s,s',s'',Q^2)=
\varepsilon_{\mu\mu\alpha\beta}q^\alpha P^\beta
\cdot\Delta^{(1)}(s,s',s'',Q^2),
$$
where
\begin{eqnarray}
\Delta^{(1)}(s,s',s'',Q^2)&&=-4\pi^5C_F
\int \frac{d^4k_2}{i(2\pi)^4} \frac{d^4k'_2}{i(2\pi)^4}
\frac{4\pi\alpha_s(t)}{m_G^2-t}\nonumber \\
\times S(1)&&\delta(m^2-k_1^2)\delta(m^2-k_2^2)\delta(m^2-k_1'^2)
\nonumber \\ \times&&\delta(m^2-k_2'^2) \delta(m^2-k_1''^2)\ .
\end{eqnarray}

Finally, the dispersion representation for the soft--hard form
factor $F_{\gamma\pi}^{SH(1)}(Q^2)$
reads
\begin{eqnarray} \label{fsh1}
&F&_{\gamma\pi}^{SH(1)}(Q^2)=2 Z_\pi \sqrt{N_c} \int
\frac{ds ds' ds''}{\pi\;\pi\;\pi}
\Psi_\pi(s)\frac{G_\gamma(s')}{s'}\frac 1{s''} \nonumber \\
&\times&\Delta^{(1)}(s,s',s'',Q^2)\theta(s_0-s)
\theta(s_0-s')\theta(s''-s_0).
\end{eqnarray}

Introducing light--cone variables $x=k_{2+}/P_+$, $\vec k_\perp =\vec
k_{2\perp}$, $x'=k'_{2+}/P_+$, $\vec k_\perp' =\vec k'_{2\perp}$ (see
also Eqs.\ (\ref{lccom}), (\ref{lcdif})), one has:
\begin{eqnarray} \label{D1lc}
&\Delta&^{(1)}(s,s',s'',Q^2)=\frac{C_F}{256\pi^3} \int
\frac{dxd^2\vec k_\perp}{x(1-x)^2}\frac{dx'd^2\vec k_\perp'}{x'(1-x')}
\nonumber \\ &\times&\frac{4\pi\alpha_s(t)}{m_G^2-t} S(1)
\delta\left(s -\frac{m^2+\vec k_\perp^2}{x (1-x )}\right)
\delta\left(s'-\frac{m^2+\vec k_\perp'^2}{x'(1-x')}\right)\nonumber \\
&\times&\delta\left(s''-\frac{m^2+(\vec k_\perp-x\vec Q)^2}{x(1-x )}
\right).
\end{eqnarray}

Substituting Eq.\ (\ref{D1lc}) into Eq.\ (\ref{fsh1}) yields the
expression of $F_{\gamma\pi}^{SH(1)}(Q^2)$ in terms of the
light--cone variables:
\begin{eqnarray} \label{fsh1f}
F_{\gamma\pi}^{SH(1)}(Q^2)&=& \frac{2 Z_\pi C_F\sqrt{N_c}}
{(16\pi^3)^2}
\int\frac{dxd^2\vec k_\perp}{x(1-x)^2}\frac{dx'd^2\vec k_\perp'}
{x'(1-x')}\nonumber\\
&\times&\Psi_\pi(s)\frac{G_\gamma(s')}{s'}\frac1{s''}
\frac{4\pi\alpha_s(t)}{m_G^2-t} S(1)
\nonumber\\&\times&\theta(s_0-s)\theta(s_0-s')\theta(s''-s_0),
\end{eqnarray}
where
\begin{eqnarray} \label{scond1}
s  =\frac{m^2+\vec k_\perp^2}{x (1-x )}\ &,&\quad
s' =\frac{m^2+(\vec k_\perp'-x'\vec Q)^2}{x'(1-x')}\ ,\nonumber\\
s''&=&\frac{m^2+(\vec k_\perp-x\vec Q)^2}{x (1-x )},
\end{eqnarray}
\begin{equation} \label{t}
t=-\frac{m^2(x'-x)^2+(x\vec k_\perp'-x'\vec k_\perp)^2}{x'x}
\end{equation}

The light--cone expressions for $a_1',a_2',b',b''$ are rather simple
(see Eqs.\ (\ref{acoefs}), (\ref{bcoefs})), and the function
$S(1)$ takes the form
\begin{eqnarray}
&&S(1)=4m\left[s'
-12\left(\frac{(\vec k_\perp\vec Q)(\vec k_\perp'\vec Q)}{Q^2}
-(\vec k_\perp \vec k_\perp' )\right)\right.
\nonumber\\
&&\left.+4\left(\frac{(\vec k_\perp'\vec Q)^2}{Q^2}-\vec k_\perp'^2
\right)-(s''-s+Q^2)\frac{(\vec k_\perp' \vec Q )}{Q^2}\right].
\end{eqnarray}

\subsection{Soft--hard term $F_{\gamma\pi}^{SH(2)}(Q^2)$}
\label{sec-sh2}
Calculation of the soft--hard form factor
$F_{\gamma\pi}^{SH(2)}(Q^2)$ looks very much like the calculation of
$F_{\gamma\pi}^{SH(1)}(Q^2)$: $F_{\gamma\pi}^{SH(2)}(Q^2)$ is
connected with the amplitude of the diagram Fig.\ \ref{fig:series}d
which can be written as the following dispersion representation
\begin{eqnarray}
&T&_{\mu\nu}^{SH(2)}(Q^2)= \nonumber \\
&2&e^2 Z_\pi \sqrt{N_c} \int
\frac{ds ds' ds''}{\pi\;\pi\;\pi}
\Psi_\pi(s)\frac{G_\gamma(s')}{s'}\frac 1{s''-m_\pi^2}\nonumber \\
&\times&D_{\mu\nu}^{(2)}(s,s',s'',Q^2)\theta(s_0-s)
\theta(s_0-s')\theta(s''-s_0).
\end{eqnarray}
Here $D_{\mu\nu}^{(2)}(s,s',s'',Q^2)$ is the spectral density of the
diagram of Fig.\ \ref{fig:series}d which reads
\begin{eqnarray}
D_{\mu\nu}^{(2)}(s,s',s'',Q^2)&&=C_F (2\pi i)^5 \frac{(-i)^3}8
\int \frac{d^4k_2}{i(2\pi)^4} \frac{d^4k'_2}{i(2\pi)^4} \nonumber \\
\times\frac{4\pi\alpha_s(t)}{m_G^2-t} Sp_{\mu\nu}^{(2)}&&
\delta(m^2-k_1^2) \delta(m^2-k_2^2) \delta(m^2-k_1'^2) \nonumber \\
\times&&\delta(m^2-k_2'^2) \delta(m^2-k_1''^2)\ , \nonumber\\
s=(k_1+k_2)^2,\; s'&&=(k_1'+k_2')^2,\; s''=(k_1''+k_2')^2.
\end{eqnarray}

$Sp_{\mu\nu}^{(2)}$ denotes the trace
\begin{eqnarray}
&&Sp_{\mu\nu}^{(2)}=Sp\left(i\gamma_5(m-\hat k_2)
\gamma_\alpha(m-\hat k_2')\gamma_\nu(m+\hat k_1')
\gamma_\mu \right.\nonumber \\
&&\hspace{3.8cm}\left.
\times(m+\hat k_1'')\gamma_\alpha(m+\hat k_1)\right).
\end{eqnarray}

A detailed calculation of the quantity $Sp_{\mu\nu}^{(2)}$ is
presented in Appendix. Isolating the factor
$\varepsilon_{\mu\mu\alpha\beta}q^\alpha P^\beta $ one finds
$$
Sp_{\mu\nu}^{(2)}=\varepsilon_{\mu\mu\alpha\beta}q^\alpha P^\beta
\cdot S(2)
$$
\begin{eqnarray}
S(2)=8m\left[ s \right. &&-(a_1'-a_2')(k_1'' P)-a_2'(k_1'k_2')
-2b''\nonumber \\ &&\left. -m^2(1+a_2')\right].
\end{eqnarray}
The expressions for $a_1',a_2',b',b''$ are given in Appendix
(Eqs.\ (\ref{kmu}), (\ref{acoefs}) and (\ref{bcoefs})).

So,
$$
D_{\mu\nu}^{(2)}(s,s',s'',Q^2)=
\varepsilon_{\mu\mu\alpha\beta}q^\alpha P^\beta
\cdot\Delta^{(2)}(s,s',s'',Q^2)
$$
\begin{eqnarray}
\Delta^{(2)}(s,s',s'',Q^2)&&=-4\pi^5C_F
\int \frac{d^4k_2}{i(2\pi)^4} \frac{d^4k'_2}{i(2\pi)^4}
\frac{4\pi\alpha_s(t)}{m_G^2-t}\nonumber \\
\times S(2)&&\delta(m^2-k_1^2)\delta(m^2-k_2^2)\delta(m^2-k_1'^2)
\nonumber \\ \times&&\delta(m^2-k_2'^2) \delta(m^2-k_1''^2)\ .
\end{eqnarray}

Finally, the soft--hard form factor $F_{\gamma\pi}^{SH(2)}(Q^2)$
reads
\begin{eqnarray} \label{fsh2}
&F&_{\gamma\pi}^{SH(2)}(Q^2)= \nonumber \\
&2& Z_\pi \sqrt{N_c} \int
\frac{ds ds' ds''}{\pi\;\pi\;\pi}
\Psi_\pi(s)\frac{G_\gamma(s')}{s'}\frac 1{s''-m_\pi^2} \nonumber \\
&\times&\Delta^{(2)}(s,s',s'',Q^2)\theta(s_0-s)
\theta(s_0-s')\theta(s''-s_0).
\end{eqnarray}

Using light--cone variables one gets
\begin{eqnarray} \label{D2lc}
&\Delta&^{(2)}(s,s',s'',Q^2)=\frac{C_F}{256\pi^3} \int
\frac{dxd^2\vec k_\perp}{x(1-x)}\frac{dx'd^2\vec k_\perp'}
{x'(1-x')^2} \nonumber \\
&\times&\frac{4\pi\alpha_s(t)}{m_G^2-t} S(2)
\delta\left(s -\frac{m^2+\vec k_\perp^2}{x (1-x )}\right)
\delta\left(s'-\frac{m^2+\vec k_\perp'^2}{x'(1-x')}\right)
\nonumber \\
&\times&\delta\left(s''-\frac{m^2+(\vec k_\perp'+x'\vec Q)^2}
{x'(1-x')}\right).
\end{eqnarray}

Substituting Eq.\ (\ref{D2lc}) into Eq.\ (\ref{fsh2}) leads to the
expression for $F_{\gamma\pi}^{SH(2)}(Q^2)$ in terms of the
light--cone variables:
\begin{eqnarray} \label{fsh2f}
F_{\gamma\pi}^{SH(2)}(Q^2)&=& \frac{2 Z_\pi C_F\sqrt{N_c}}
{(16\pi^3)^2}
\int\frac{dxd^2\vec k_\perp}{x(1-x)}\frac{dx'd^2\vec k_\perp'}
{x'(1-x')^2}\nonumber\\
&\times&\Psi_\pi(s)\frac{G_\gamma(s')}{s'}\frac1{s''-m_\pi^2}
\frac{4\pi\alpha_s(t)}{m_G^2-t} S(2)
\nonumber\\&\times&\theta(s_0-s)\theta(s_0-s')\theta(s''-s_0),
\end{eqnarray}
where
\begin{equation} \label{scond2}
s  =\frac{m^2+\vec k_\perp^2  }{x (1-x )},
s' =\frac{m^2+(\vec k_\perp'-x'\vec Q )^2}{x'(1-x')},
s''=\frac{m^2+\vec k_\perp'^2 }{x'(1-x')}.
\end{equation}

Using the light--cone expressions for $a_1',a_2',b',b''$ (see Eqs.\
(\ref{acoefs}), (\ref{bcoefs})), one gets the following result for
$S(2)$:
\begin{eqnarray}
&&S(2)=8m\left[s+\frac{(\vec k_\perp' \vec Q )}{Q^2}(k_1''P)
-\left(x'+\frac{(\vec k_\perp' \vec Q )}{Q^2}\right)(k_1'k_2')
\right.\nonumber\\ &&\left.
-2\left(\frac{(\vec k_\perp' \vec Q )^2}{Q^2}-\vec k_\perp'^2 \right)
-m^2\left(1+x'+\frac{(\vec k_\perp' \vec Q )}{Q^2}\right)
\right].
\end{eqnarray}

\subsection{Calculation results} \label{sec-respi}

Calculations of form factors at low and intermediate $Q^2$ include
masses which are related to the soft physics: constituent quark mass
and effective gluon mass. We use standard constituent quark mass
$m=0.35\;GeV$ while for effective gluon mass, $m_G$, we consider,
just as in Ref. \cite{amn}, the following two variants (masses are
given in $GeV$):
\begin{eqnarray}\label{mg}
(i)&&\quad m_G=0\ ,\nonumber\\
(ii)&&\quad m_G=\left\{\begin{array}{cr}
      0.7\left(1+\frac t{0.5}\right) &,\ |t|<0.5\;GeV^2\ , \\
      0   &,\ |t|>0.5\;GeV^2\ .\end{array}\right.
\end{eqnarray}

\begin{figure}
\centerline{\epsfig{file=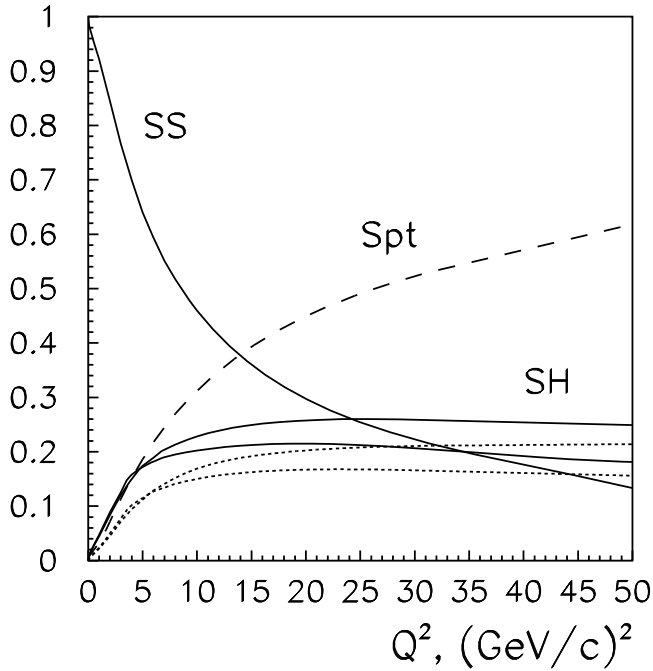,width=9.5cm}}
\caption{Relative contributions of different terms to the form
factor: the ratios of different terms contributing to the whole
form factor are shown. Four variants for the $SH$--term correspond to:
(1) $m_G=0$, $\alpha_s(t)$ -- the upper solid curve;
(2) $m_G=0$, $\alpha_s\left(\frac{Q^2}4\right)$ -- the lower solid
curve;
(3) $m_G=m_G(t)$, $\alpha_s(t)$ -- the upper dotted curve;
(4) $m_G=m_G(t)$, $\alpha_s\left(\frac{Q^2}4\right)$ -- the lower
dotted curve.}
\label{fig:rats}
\end{figure}

For $\alpha_s(t)$ we use  two parametrization:
\begin{eqnarray}\label{varalp}
(i)&&\quad \alpha_s(t)=\left\{\begin{array}{cr}
 const &,\ |t|<1\;GeV^2\ ,\\
  \frac{4\pi}9 ln^{-1}\left(\frac{|t|}{\Lambda^2}\right)
  &,\ |t|>1\;GeV^2\ ,\end{array}\right.\nonumber\\
(ii)&&\quad \alpha_s(t)=\alpha_s\left(\frac{Q^2}4\right).
\end{eqnarray}
The first variant corresponds to a frozen $\alpha_s(t)$ at low $t$:
it is the same variant as in Eq.\ (\ref{fq}). In the second variant we
take $\alpha_s(t)$ in the middle point, $|t|\to Q^2/4$. So, Eqs.\
(\ref{mg}) and (\ref{varalp}) provide four possible variants for the
soft--hard form factors. Corresponding results for soft--hard term are
shown in Fig.\ \ref{fig:rats}. The main conclusion coming from the
study of these variants is that a concrete choice of $m_G$ and
$\alpha_S$ in the soft region does not much influence the result.

\begin{figure}
\centerline{\epsfig{file=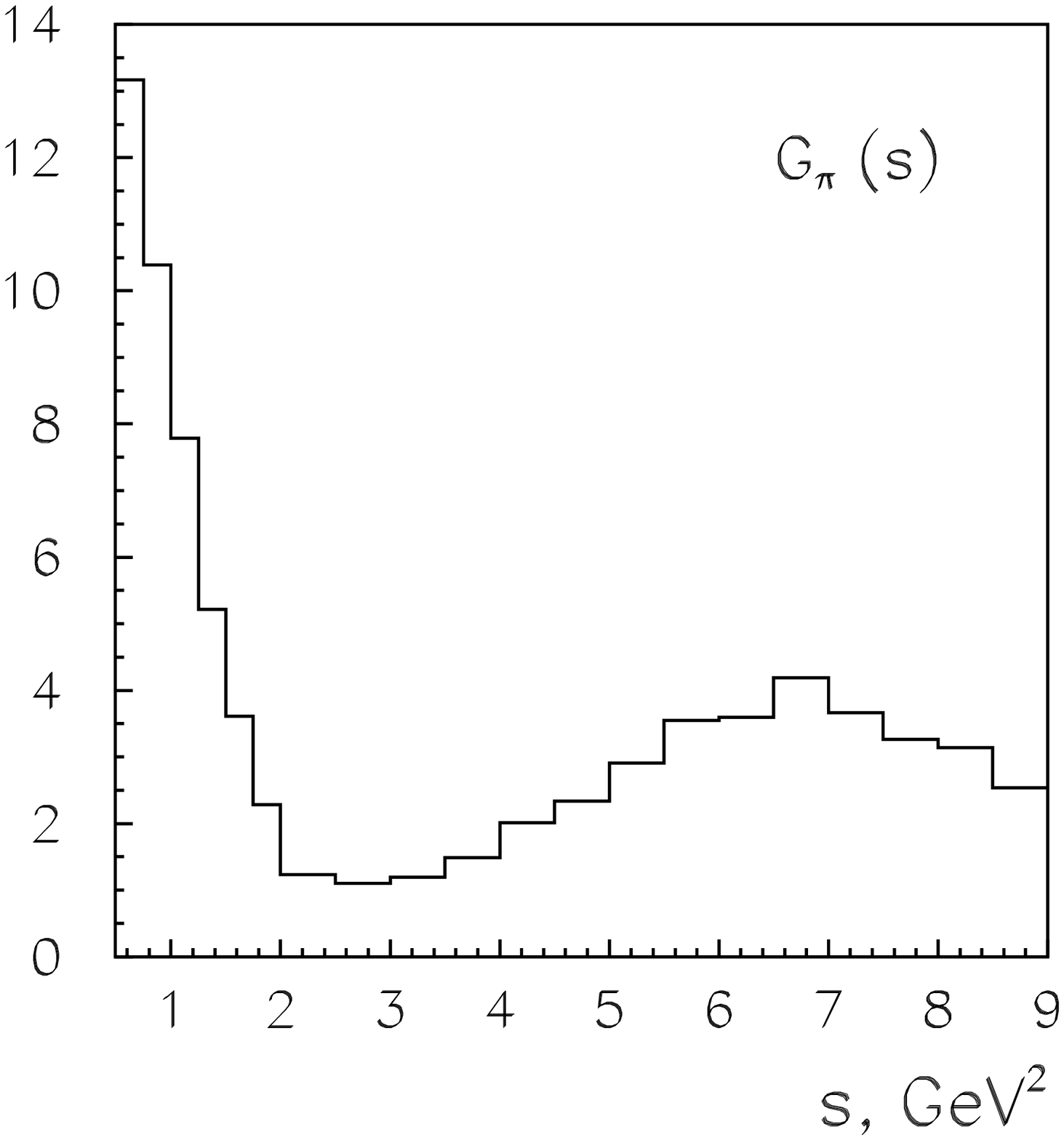,width=8.5cm}}
\centerline{\epsfig{file=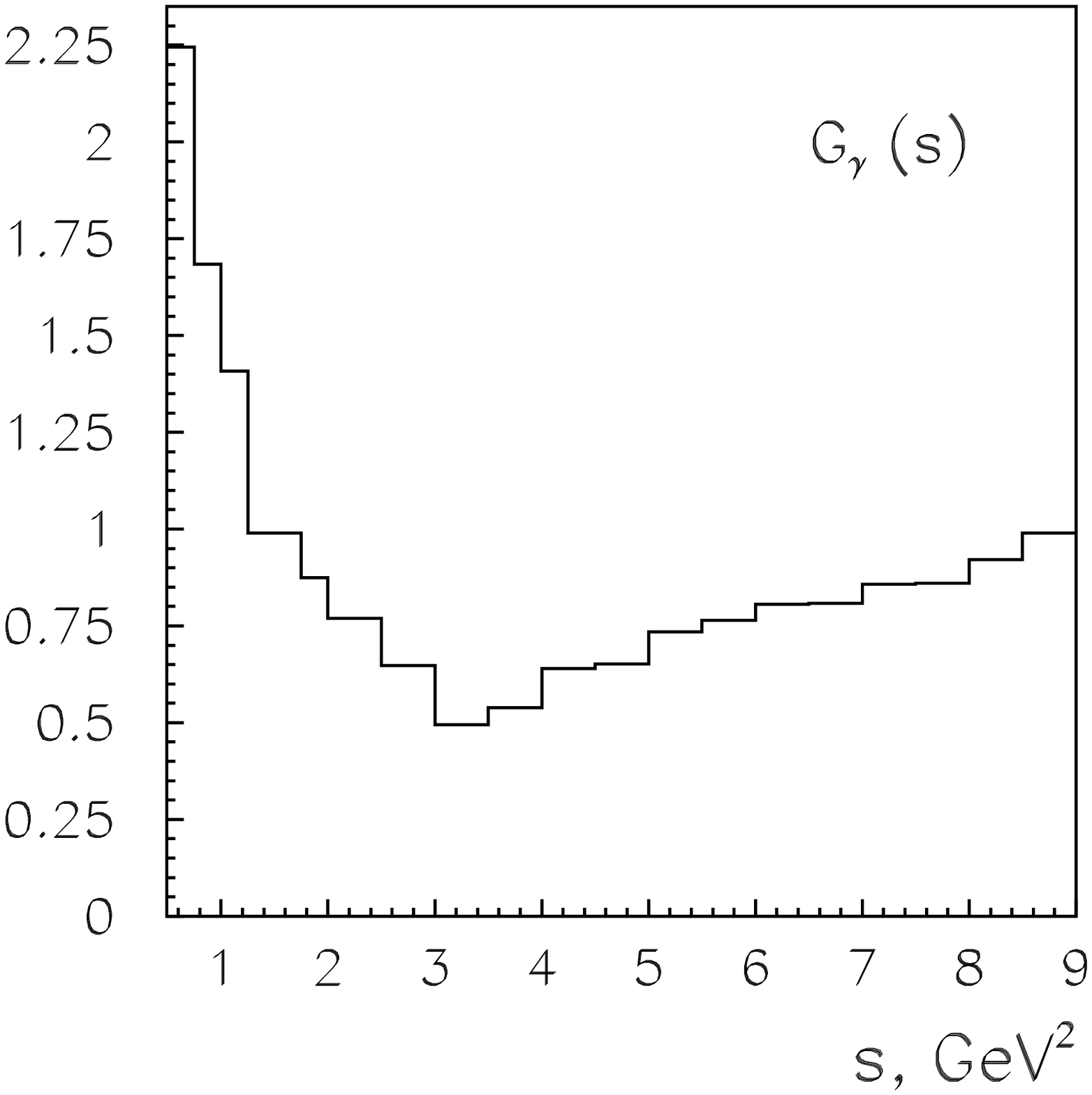,width=8.5cm}}
\caption{ $q\bar q$--distribution function for pion (a) and for
photon (b).}
\label{fig:wf}
\end{figure}

\begin{figure}
\centerline{\epsfig{file=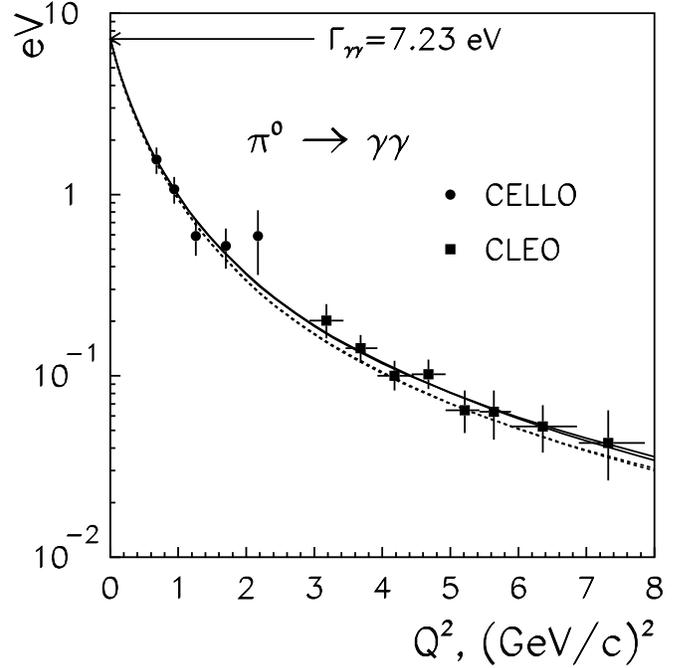,width=9.5cm}}
\caption{$\gamma\pi^0$ transition form factor: the quantity
$\frac\pi4\alpha^2m_\pi^3F_{\gamma\pi}^2(Q^2)$ is shown. Different
curves correspond to different sets of parameters in the $SH$-term;
the curve notation is the same as in Fig. 2. Experimental data are
taken from Ref. [2].}
\label{fig:ffpi}
\end{figure}

Much more important is a choice of the soft photon wave function.
There are two constraints on the soft photon wave function: the
normalization condition (\ref{gnorm}) which guarantees the correct
decay widht $\pi^0\to \gamma\gamma$, and the sewing condition of Eq.\
(\ref{sewing}). In addition, we impose the third constraint in the
spirit of vector dominance model and $SU(6)$--symmetry: the
slope of $G_\gamma(s)$ in the region $4m^2<s<1.5\;GeV^2$ is the same
as that of $\pi(s)$. Using these constraints we reconstruct the photon
$q\bar q$--distribution function $G_\gamma(s)$ which is shown in Fig.
\ref{fig:wf}, together with $G_\pi(s)$ for the comparison. The
reconstructed distribution function has a dip in the region $s \sim
2-4\; GeV^2$ similar to that of $G_\pi(s)$. In  Ref. \cite{amn} the
$q\bar q$--distribution in the pion is referred as quasizone one. The
reconstructed $G_\gamma(s)$ behaviour provides an argument that the
quasizone structure in the $q\bar q$--distributions is of a common
nature. The quantity $\frac\pi4\alpha^2m_\pi^3F_{\gamma\pi}^2(Q^2)$
with the reconstructed $G_\gamma(s)$ is presented in Fig.\ \ref{fig:ffpi}.

\begin{figure}
\centerline{\epsfig{file=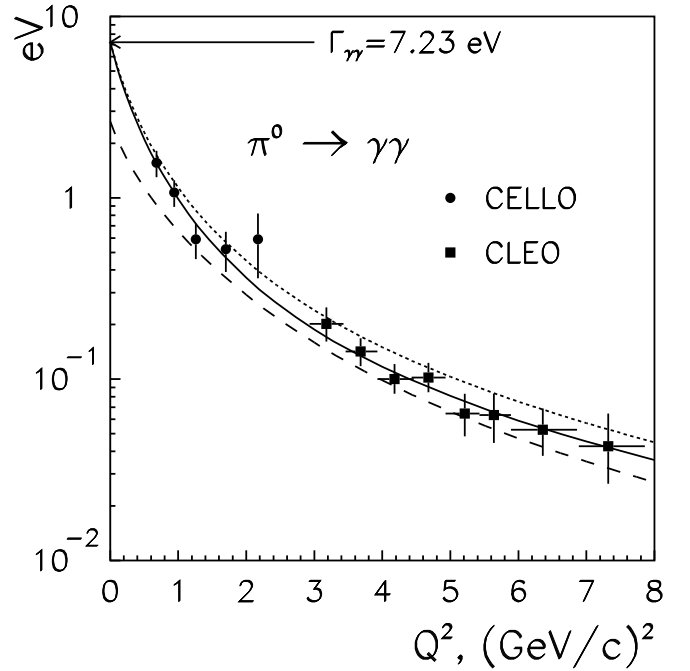,width=9.5cm}}
\caption{The quantity $\frac\pi4\alpha^2m_\pi^3F_{\gamma\pi}^2(Q^2)$
for the variant (1) of Eq. (57) -- dotted line, and for the variant
(2) -- dashed line. The solig line is the same as in Fig. 4.}
\label{fig:ffdif}
\end{figure}

To check  the form factor  sensitivity to the $q\bar q$--distribution
function we have calculated $F_{\gamma\pi}(Q^2)$ with alternative
choices of $G_\gamma(s)$:
\begin{eqnarray}
(1)&&\quad G_\gamma^{(1)}(s)=\left\{\begin{array}{cr}
      0.98G_\gamma(s) &,\ s\le 1.5\;GeV^2\ , \\
      1   &,\ s>1.5\;GeV^2\ ,\end{array}\right.,\nonumber\\
(2)&&\quad G_\gamma^{(2)}(s)=1\ .
\end{eqnarray}
The results are shown in Fig.\ \ref{fig:ffdif}. The variant (1)
corresponds to the low--$s$ $q\bar q$--distribution similar to
$G_\pi(s)$ but without a dip at $s\sim 3\; GeV^2$ (the factor 0.98 is
due to renormalization of the $q\bar q$--distribution to have
$\Gamma_{\gamma\gamma}=2.23\;eV$). The absence of a dip in the photon
$q\bar q$--distribution results in raising the calculated curve at
large $Q^2$. The variant (2), which corresponds to the point-like
vertex $q\bar q\to photon$ at low $s$, represents neither
$\Gamma_{\gamma\gamma}$ nor low--$Q^2$ form factor correctly.

\section{$\gamma\eta$ and $\gamma\eta'$ transition form factors}
\label{sec-etaff}

In our calculations of the $\gamma\eta$ and $\gamma\eta'$ transition
form factors we assume, in the spirit of the quark model, the
universality of  soft wave functions of the $0^-$--nonet. This
universality is usually formulated in the $r$--representation, or in
terms of the relative quark momenta. So, let us rewrite the pion wave
function $\Psi_\pi(s)$ (see Eq.\ (\ref{psipi})) as a function of $\vec
k$ which is connected with the energy squared by $\vec
k^2=(s-4m^2)/4$.

For the momentum representation of the pion wave function we use the
following form
\begin{equation} \label{psik}
\Psi_\pi(s)=\psi_\pi(\vec k^2)=\frac{g(\vec k^2)}{\vec k^2+k_0^2}
\ ,\end{equation}
where according to Ref.\ \cite{amn} $k_0^2=0.1176$ GeV$^2$.

The pseudoscalar mesons $\eta$ and $\eta'$ are mixtures of the
non-strange and strange quarks, $n\bar n=(u\bar u+d\bar d)/\sqrt 2$
and $s\bar s$: $\eta =n\bar n\; Cos\theta- s\bar s\; Sin\theta$,
$\eta'=n\bar n\; Sin\theta+ s\bar s\; Cos\theta$. Respectively, the
wave functions of the $\eta$ and $\eta'$ mesons are described by the
two components
\begin{eqnarray}
\Psi_\eta   &=&Cos\theta\;\psi_{n\bar n}(\vec k^2)
             - Sin\theta\;\psi_{s\bar s}(\vec k^2)
\ ,\nonumber \\
\Psi_{\eta'}&=&Sin\theta\;\psi_{n\bar n}(\vec k^2)
             + Cos\theta\;\psi_{s\bar s}(\vec k^2)\ .
\end{eqnarray}

The universality of the pseudoscalar meson wave function means that
the function
\begin{equation}
\psi_{n\bar n}(\vec k^2)=\frac{g(\vec k^2)}{\vec k^2+k_0^2}\ ,
\end{equation}
normalized as follows
\begin{equation}
\frac 2{\pi^2}\int\limits_{0}^{(s_0-4m^2)/4}d\vec k^2\;
k\sqrt{\vec k^2+m^2}\psi_{n\bar n}^2(\vec k^2)=1,
\end{equation}
is just the same function as for the $\pi$--meson.

For the $s\bar s$--component we should take into account
strange/non--strange quark mass difference, hence
\begin{equation}
\psi_{s\bar s}(\vec k^2)=N\frac{g(\vec k^2)}
{\vec k^2+k_0^2+\Delta^2} \ ,
\end{equation}
where $\Delta^2=m_s^2-m^2$, $m_s=0.5\; GeV$ and $\vec
k^2=(s-4m_s^2)/4$. Numerical factor $N$ corresponds to a
renormalization of $\psi_{s\bar s}(\vec k^2)$ after introducing
$\Delta^2$. The factor $N$ for the strange component of the wave
function is determined by the normalization condition
\begin{equation}
\frac 2{\pi^2}\int\limits_{0}^{(s_0-4m^2)/4}d\vec k^2\;
k\sqrt{\vec k^2+m_s^2}\psi_{s\bar s}^2(\vec k^2)=1\ .
\end{equation}

Similarly, for the momentum representation of the soft photon wave
function we use the expressions
\begin{equation} \label{gamk}
\Psi_\gamma(s)\equiv\psi_{\gamma\to n\bar n}(\vec k^2)=
\frac{g_\gamma(\vec k^2)}{\vec k^2+m^2}
\end{equation}
and
\begin{equation}
\psi_{\gamma\to s\bar s}(\vec k^2)=
\frac{g_\gamma(\vec k^2)}{\vec k^2+m^2+\Delta^2}.
\end{equation}

Thus, the $n\bar n$ and $s\bar s$ components of the meson and
soft photon wave function being fixed, one can proceed with
calculations of the transition $\gamma\eta$ and $\gamma\eta'$ form
factors, which are determined as
\begin{eqnarray}
F_{\gamma\eta} (Q^2)&=&Cos\theta\;F_{n\bar n}(Q^2)
                     - Sin\theta\;F_{s\bar s}(Q^2)\ ,\nonumber \\
F_{\gamma\eta'}(Q^2)&=&Sin\theta\;F_{n\bar n}(Q^2)
                     + Cos\theta\;F_{s\bar s}(Q^2)\ .
\end{eqnarray}

In the next two subsections we present formulae for $F_{n\bar n}(Q^2)$
and $F_{s\bar s}(Q^2)$.

\subsection{The $n\bar n$ contributions to the form factor}

These contributions can be obtained from the corresponding terms
in $F_{\gamma\pi}(Q^2)$ by
replacing the charge factor and wave functions.
The final results are listed below.

Soft--soft term $F_{n\bar n}^{SS}(Q^2)$:
\begin{eqnarray} \label{ssnn}
F_{n\bar n}^{SS}(Q^2)&=&
2Z_{n\bar n} f_q(Q^2)\sqrt{N_c}\frac m{4\pi^5}
\int \frac{dxd^2\vec k_\perp }{x(1-x)^2} \nonumber \\
&\times&\psi_{n\bar n}(\vec k^2)\theta(s_0-s)
\frac{g_\gamma(\vec k'^2)}{s'}\theta(s_0-s'),
\end{eqnarray}
\begin{equation}
Z_{n\bar n}=\frac{e_u^2+e_d^2}{\sqrt 2},\quad
\vec k^2=\frac s4-m^2,\quad
\vec k'^2=\frac {s'}4-m^2,
\end{equation}
\begin{eqnarray}
F_{n\bar n}^{SS}(0)&&=Z_{n\bar n}\frac m{2\pi}\nonumber \\
&&\times\int\limits_{4m^2}^{s_0}
\frac{ds}{\pi}\psi_{n\bar n}(\vec k^2)\frac{g_\gamma(\vec k^2)}s
ln\frac{1+\sqrt{1-4m^2/s}}{1-\sqrt{1-4m^2/s}}.
\end{eqnarray}
$s$ and $s'$ are given by Eq.\ (\ref{slc}).

Soft--point term $F_{n\bar n}^{Spt}(Q^2)$:
\begin{eqnarray}\label{spnn}
F_{n\bar n}^{SPt}(Q^2)&=&
2Z_{n\bar n}f_q(Q^2)\sqrt{N_c}\frac m{4\pi^5}
\int \frac{dxd^2\vec k_\perp }{x(1-x)^2} \nonumber \\&\times&
\psi_{n\bar n}(\vec k^2)\frac 1{s'}\theta(s_0-s)\theta(s'-s_0).
\end{eqnarray}

Soft--hard term $F_{n\bar n}^{SH(1)}(Q^2)$:
\begin{eqnarray} \label{sh1nn}
F_{n\bar n}^{SH(1)}(Q^2)&=& \frac{2 Z_{n\bar n} C_F\sqrt{N_c}}
{(16\pi^3)^2}
\int \frac{dxd^2\vec k_\perp}{x(1-x)^2}\frac{dx'd^2\vec k_\perp'}
{x'(1-x')}\nonumber\\
&\times&\psi_{n\bar n}(\vec k^2)\frac{g_\gamma(\vec k'^2)}{s'}
\frac1{s''}\frac{4\pi\alpha_s(t)}{m_G^2-t} S(1)
\nonumber\\&\times&\theta(s_0-s)\theta(s_0-s')\theta(s''-s_0),
\end{eqnarray}
$s$, $s'$ and $s''$ are given by Eq.\ (\ref{scond1}).

Soft--hard term $F_{n\bar n}^{SH(2)}(Q^2)$:
\begin{eqnarray} \label{sh2nn}
F_{n\bar n}^{SH(2)}(Q^2)&=& \frac{2 Z_{n\bar n} C_F\sqrt{N_c}}
{(16\pi^3)^2}
\int\frac{dxd^2\vec k_\perp}{x(1-x)}\frac{dx'd^2\vec k_\perp'}
{x'(1-x')^2}\nonumber\\
&\times&\psi_{n\bar n}(\vec k^2)\frac{g_\gamma(\vec k'^2)}{s'}
\frac1{s''-m_{\eta,\eta'}^2}\frac{4\pi\alpha_s(t)}{m_G^2-t} S(2)
\nonumber\\&\times&\theta(s_0-s)\theta(s_0-s')\theta(s''-s_0),
\end{eqnarray}
$s$, $s'$ and $s''$ are given by Eq.\ (\ref{scond2}).

\subsection{The $s\bar s$ contributions to the form factor}

These expressions are obtained from the corresponding terms in
$F_{\gamma\pi}(Q^2)$ by
replacing the charge factor, wave functions and quark masses.
The results have the following form.

Soft--soft term $F_{s\bar s}^{SS}(Q^2)$:
\begin{eqnarray} \label{ssss}
F_{s\bar s}^{SS}(Q^2)&=&
2Z_{s\bar s} f_q(Q^2)\sqrt{N_c}\frac {m_s}{4\pi^5}
\int \frac{dxd^2\vec k_\perp }{x(1-x)^2} \nonumber \\
&\times&\psi_{s\bar s}(\vec k^2)\theta(s_0-s)
\frac{g_\gamma(\vec k'^2)}{s'}\theta(s_0-s')\ ,
\end{eqnarray}
\begin{eqnarray}
&&Z_{s\bar s}=e_s^2,\quad
\vec k^2=\frac s4-m_s^2,\quad
\vec k'^2=\frac {s'}4-m_s^2\ , \\
&&s=\frac{m_s^2+\vec k_\perp^2 }{x(1-x )}, \;
s'=\frac{m_s^2+(\vec k_\perp -x\vec Q )^2}{x(1-x)}.
\end{eqnarray}
\begin{eqnarray}
F_{s\bar s}^{SS}&&(0)=Z_{s\bar s}\frac{m_s}{2\pi}\nonumber \\
&&\times\int\limits_{4m_s^2}^{s_0+4\Delta^2}
\frac{ds}\pi\psi_{s\bar s}(\vec k^2)\frac{g_\gamma(\vec k^2)}s
ln\frac{1+\sqrt{1-4m_s^2/s}}{1-\sqrt{1-4m_s^2/s}}
\end{eqnarray}

Soft--point term $F_{s\bar s}^{Spt}(Q^2)$:
\begin{eqnarray}\label{spss}
F_{s\bar s}^{SPt}(Q^2)&=&
2Z_{s\bar s}f_q(Q^2)\sqrt{N_c}\frac {m_s}{4\pi^5}
\int \frac{dxd^2\vec k_\perp }{x(1-x)^2} \nonumber \\ &\times&
\psi_{s\bar s}(\vec k^2)\frac 1{s'}\theta(s_0-s)\theta(s'-s_0).
\end{eqnarray}

Soft--hard term $F_{s\bar s}^{SH(1)}(Q^2)$:
\begin{eqnarray} \label{sh1ss}
F_{s\bar s}^{SH(1)}(Q^2)&=& \frac{2 Z_{s\bar s} C_F\sqrt{N_c}}
{(16\pi^3)^2}
\int \frac{dxd^2\vec k_\perp }{x(1-x)^2} \frac{dx'd^2\vec k_\perp' }
{x'(1-x')}\nonumber\\
&\times&\psi_{s\bar s}(\vec k^2)\frac{g_\gamma(\vec k'^2)}{s'}
\frac1{s''}\frac{4\pi\alpha_s(t)}{m_G^2-t} S(1)
\nonumber\\&\times&\theta(s_0-s)\theta(s_0-s')\theta(s''-s_0),
\end{eqnarray}
\begin{eqnarray}
s  =\frac{m_s^2+\vec k_\perp^2}{x (1-x )}\ &,&\quad
s' =\frac{m_s^2+(\vec k_\perp'-x'\vec Q)^2}{x'(1-x')}\ ,\nonumber\\
s''&=&\frac{m_s^2+(\vec k_\perp-x\vec Q)^2}{x (1-x )},
\end{eqnarray}
\begin{equation}
t=-\frac{m_s^2(x'-x)^2+(x\vec k_\perp'-x'\vec k_\perp)^2}{x'x}
\end{equation}

Soft--hard term $F_{s\bar s}^{SH(2)}(Q^2)$:
\begin{eqnarray} \label{sh2ss}
F_{s\bar s}^{SH(2)}(Q^2)&=& \frac{2 Z_{s\bar s} C_F\sqrt{N_c}}
{(16\pi^3)^2}
\int \frac{dxd^2\vec k_\perp }{x(1-x)} \frac{dx'd^2\vec k_\perp' }
{x'(1-x')^2}\nonumber\\
&\times&\psi_{s\bar s}(\vec k^2)\frac{g_\gamma(\vec k'^2)}{s'}
\frac1{s''-m_{\eta,\eta'}^2}\frac{4\pi\alpha_s(t)}{m_G^2-t} S(2)
\nonumber\\&\times&\theta(s_0-s)\theta(s_0-s')\theta(s''-s_0),
\end{eqnarray}
\begin{equation}
s  =\frac{m_s^2+\vec k_\perp^2  }{x (1-x )},
s' =\frac{m_s^2+(\vec k_\perp'-x'\vec Q )^2}{x'(1-x')},
s''=\frac{m_s^2+\vec k_\perp'^2 }{x'(1-x')}.
\end{equation}

\subsection{Calculation results for the $\gamma\eta$ and $\gamma\eta'$
transition form factors}
\label{sec-reseta}

\begin{figure}
\centerline{\epsfig{file=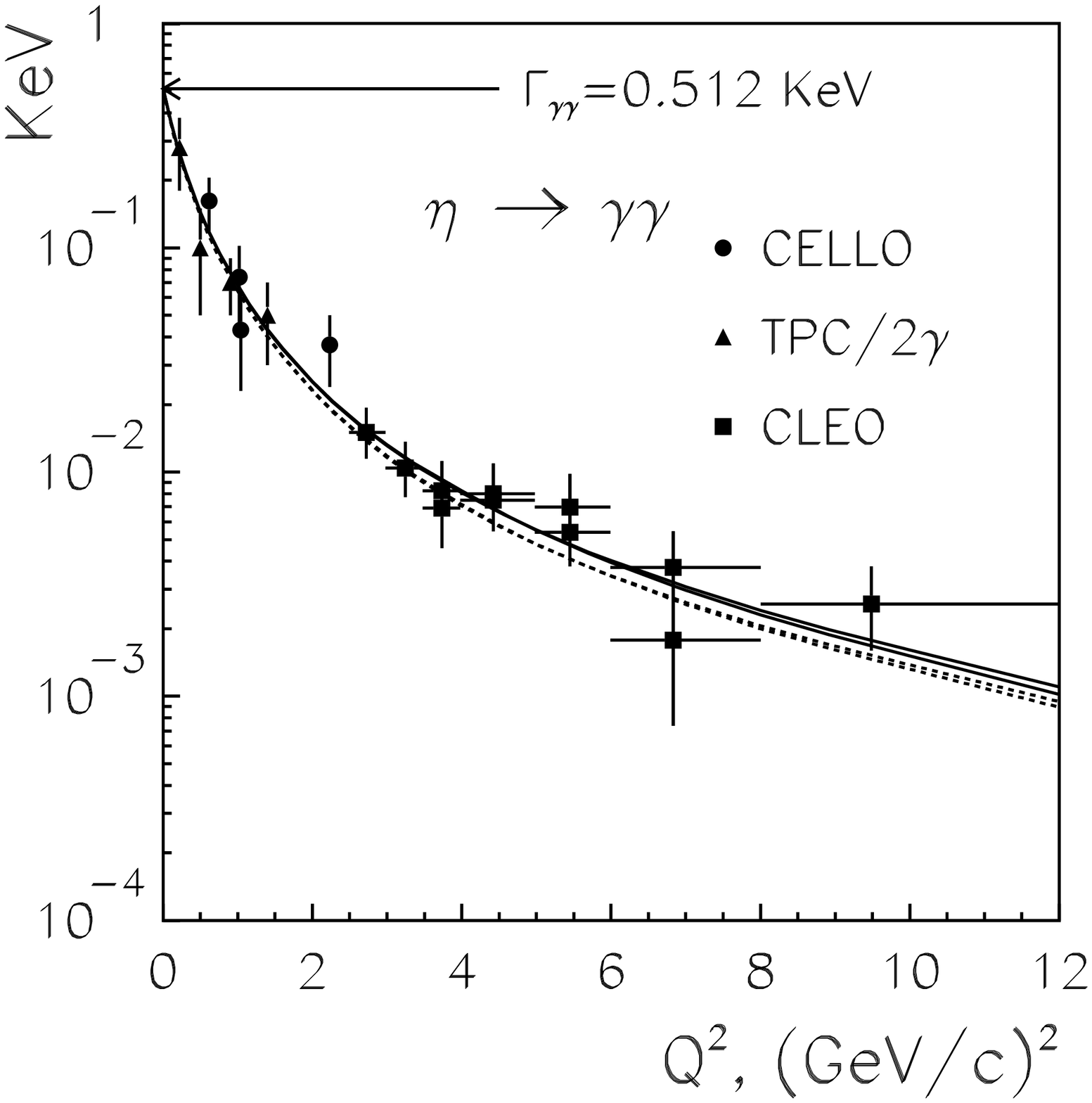,width=9.5cm}}
\centerline{\epsfig{file=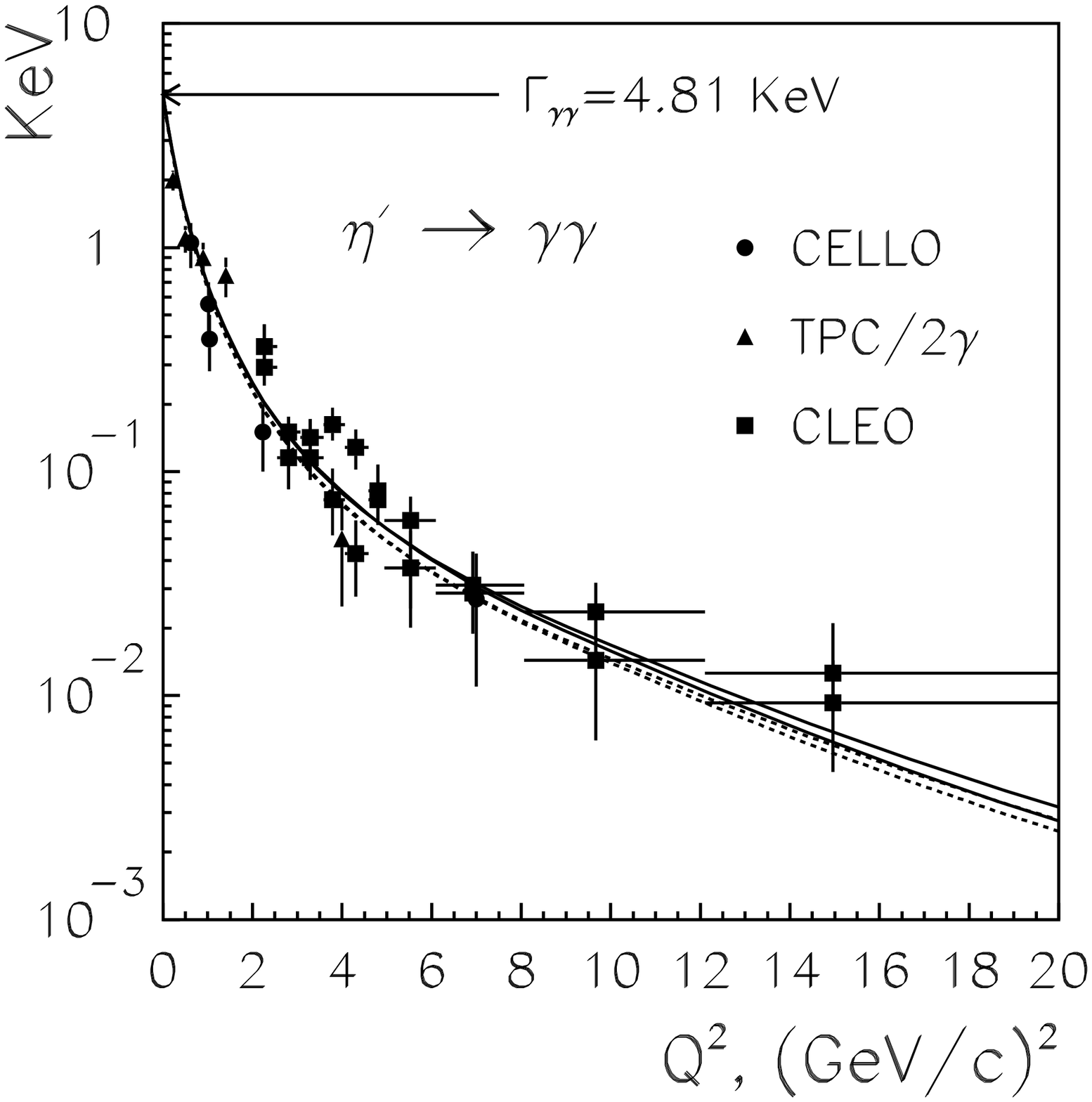,width=9.5cm}}
\caption{$Q^2$-dependence for the quantities
$\frac\pi4\alpha^2m_\eta^3F_{\gamma\eta}^2(Q^2)$ and
$\frac\pi4\alpha^2m_\eta'^3F_{\gamma\eta'}^2(Q^2)$. Different curves
correspond to different sets of the parameters in the $SH$-term (the
curve notation is the same as in Fig. 2). Experimental data are taken
from Ref. [2].}
\label{fig:ffeta}
\end{figure}

Calculation of the $\gamma\eta$ and $\gamma\eta'$ transition form
factors does not require any additional parameter, for all unknown
quantities are fixed by the $\gamma\pi$ case. We use only the
universality of wave functions of the ground--state pseudoscalar meson
nonet; we put also $Sin\theta=0.61$ \cite{pdg}. The results show an
excellent agreement with data both in shapes of the $Q^2$ dependence
of  form factors (see Fig.\ \ref{fig:ffeta}) and in absolute values of
$F_{\gamma\eta}(0)$ and $F_{\gamma\eta'}(0)$: experimental data for
the partial decay widths are $\Gamma_{\eta\to\gamma\gamma}=0.514\pm
0.052\;KeV$ \cite{pdg} and $\Gamma_{\eta'\to\gamma\gamma}=4.57\pm
0.69\;KeV$ \cite{pdg} while our calculation gives
$\Gamma_{\eta\to\gamma\gamma}=0.512\;KeV$ and
$\Gamma_{\eta'\to\gamma\gamma}=4.81\;KeV$. The universal wave
functions are presented in Table \ref{table:wf}.

Within the hypothesis about universality of pseuduscalar meson wave
functions, we may estimate the value of gluonic (or glueball)
components in $\eta$ and $\eta'$. With $\eta=C_1 q\bar q +C_2 gg$ and
$\eta'=C'_1 q\bar q +C'_2 gg$, we obtain the following constraints for
probabilities of the gluonic components in $\eta$ and $\eta'$:
$C_2^2\le 0.1$ and $C_2'^2\le 0.2$.

\section{Conclusion} \label{sec-con}

We have analyzed the $\gamma\pi^0$, $\gamma\eta$, and $\gamma\eta'$
transition form factors at low and moderately high $Q^2$ in the
framework of the method developed in Ref. \cite{amn} which allowes us
to cover the range of momenta from the soft to PQCD physics.
Calculation results back up the efficiency of the proposed method.

This consideration support the results of Ref. \cite{amn} where the
pion wave function has been reconstructed. Here we have reconstructed the
soft $q\bar q$--distribution of the photon: it has been found to be
similar to the quark distribution in the pion, that is quite natural
in the framework of the vector dominance model for soft $q\bar q$.

The calculated $\gamma\eta$ and $\gamma\eta'$ transition form factors
are in a perfect agreement with the experimental data, once we assume
the universality of the soft wave functions of the ground--state
pseudoscalar mesons, the members of the lightest nonet. This
assumptions looks very natural and inherent to the conventional quark
model.

The transition photon--meson form factors at low $Q^2$ were calculated
in Refs. \cite{jaus},\cite{muenz} using the diagram of Fig.
\ref{fig:series}b with  point--like quark--photon vertex. Such a
treatment of low--$Q^2$ photon--meson form factor has been criticized
in Ref. \cite{anselm}. Our consideration also does not support this
picture:  whereas the hard photon can be treated as a standard QED
point-like particle, the soft photon structure is nontrivial and is
very much like the structure of other ground state mesons. This
hadron-like structure of the soft photon is crucial for the
description of the photon--meson transition form factors at low $Q^2$.

In the case when both of the photon virtualities are nonzero, our
approach recovers at large $Q^2$ the $1/Q^2$--behavior as PQCD does,
and not $1/Q_1^2Q_2^2$ as might be expected from the naive application
of the vector meson dominance. The vector meson dominance reveals
itself more in the hadron-like structure of the soft photon, and not
in a naive $1/Q_1^2$ behavior of the form factors.

The study of transition form factors in the region of a few GeV$^2$
has been considered in some papers (see Ref. \cite{krolraul} and
references therein) as a possibility to discriminate between the pion
distribution amplitudes of the Chernyak-Zhitnitsky type and the
asymptotic one of Eq.\ (\ref{phias}): it was assumed that the form
factor in this region can be described by the leading PQCD term which
is considered within  modified hard scattering picture. Our analysis
shows that such an approach is not well--justified in the region of a
few GeV$^2$: the contributiuon of the soft region is far from being
small. The detailed quantitative analysises performed in Ref.
\cite{amn} and in this paper as well as recent QCD sum rule results
\cite{radrus} show that the distribution amplitude is numerically very
close to the asymptotic form of Eq.\ (\ref{phias}) and disregard the
double--humped distribution amplitudes of the Chernyak--Zhitnitsky
type.

\section{acknowledgements}
We thank L.G.Dakhno for useful discussions and remarks. We also
garateful to International Science Foundation for the financial
support under the grant R10300.

\appendix

\section*{Trace calculations in the soft--hard contribution}

\begin{figure}
\centerline{\epsfig{file=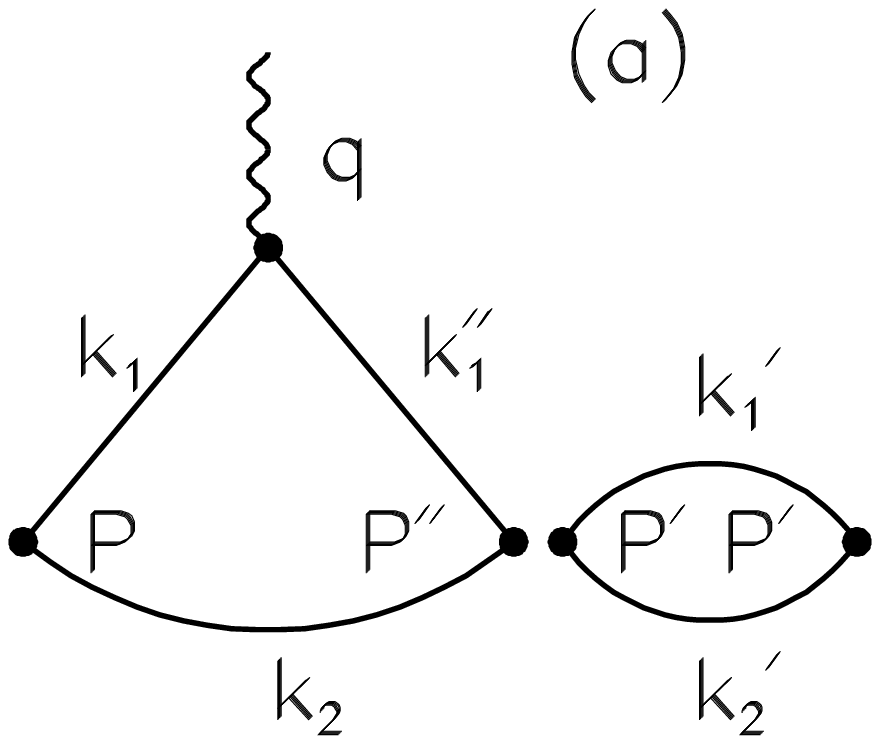,width=6cm}}
\vspace{0.5cm}
\centerline{\epsfig{file=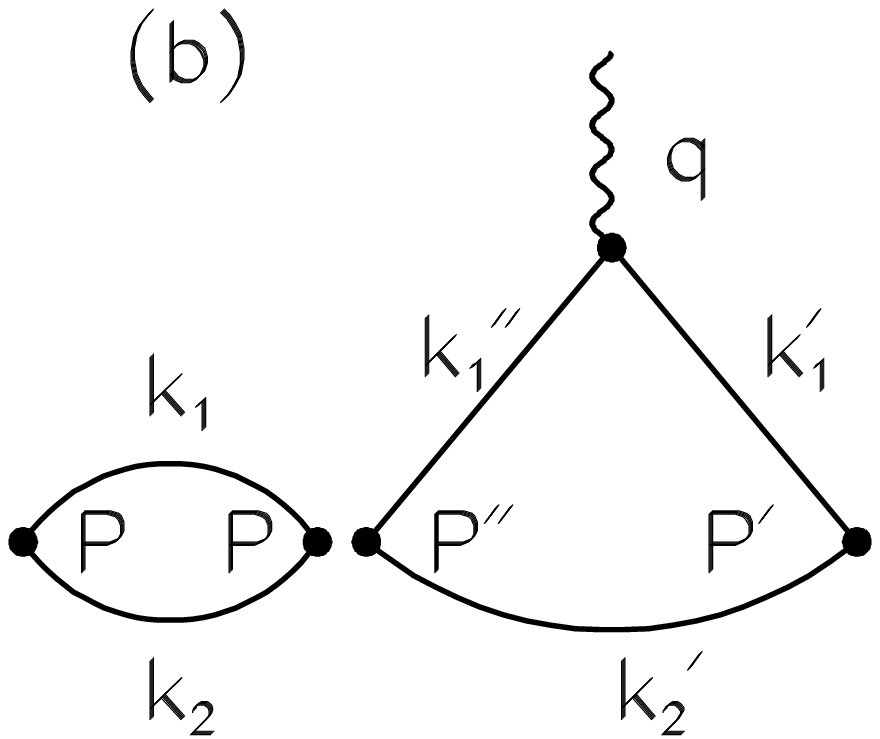,width=6cm}}
\caption{Momentum notations in  soft--hard terms
$F^{SH(1)}$ (a) and $F^{SH(2)}$ (b).}
\label{fig:loops}
\end{figure}

Soft--hard contribution is described by the two-loop diagram, and the
dispersion relation technique prescribes that the total momenta
squared of the $q\bar q$ pair should be taken different for each loop,
and then the dispersion integration over these values should be
performed. So, we must first make the Fierz rearrangements to obtain
trace calculations related to different loops. Namely, we group the
expressions as follows:
\begin{eqnarray} \label{sp1}
&&Sp\left({i\gamma_5(m-\hat k_2)
\gamma_\alpha(m-\hat k_2')\gamma_\nu(m+\hat k_1')}
\gamma_\alpha \right. \nonumber\\ &&\hspace{3cm}\left. \times
(m+\hat k_1'')\gamma_\mu(m+\hat k_1)\right)=Sp_{\mu\nu}^{(1)}=
\nonumber\\ &&\sum\limits_{i=S,V,T,A,P}C_i
Sp \left( (m+\hat k''_1)\gamma_\mu(m+\hat k_1)
i\gamma_5(m-\hat k_2)O_i\right)\nonumber \\
&&\hspace{1.5cm}\times
Sp \left({O_i(m-\hat k'_2)\gamma_\nu(m+\hat k'_1)}\right),
\end{eqnarray}
\begin{eqnarray} \label{sp2}
&&Sp\left(i\gamma_5(m-\hat k_2)
\gamma_\alpha(m-\hat k_2')\gamma_\nu(m+\hat k_1')
\gamma_\mu \right.\nonumber \\ &&\hspace{3cm}\left. \times
(m+\hat k_1'')\gamma_\alpha(m+\hat k_1)\right)=Sp_{\mu\nu}^{(2)}=
\nonumber \\ &&\sum\limits_{i=S,V,T,A,P}C_i
Sp \left({(m+\hat k_1)i\gamma_5 (m-\hat k_2)O_i}\right)\nonumber \\
&&\hspace{1cm}\times
Sp \left({O_i(m-\hat k'_2)\gamma_\nu (m+\hat k'_1)
\gamma_\mu(m+\hat k''_1)}\right),
\end{eqnarray}
with $C_S=1, C_V=-\frac12, C_T=0, C_A=\frac12, C_P =-1$. Each term
($i=S,V,T,A,P$) in Eq.\ (\ref{sp1}, \ref{sp2}) is represented as a
product of two factors, related to two different loops (see Fig.
\ref{fig:loops}). In Eq.\ (\ref{sp1}) the $S-,V-,A-$terms are nonzero
while in Eq.\ (\ref{sp2}) the $P-,A-$terms.

In calculations of the diagarams of Fig.\ \ref{fig:loops}a,b we use
the following relations,
\begin{eqnarray} \label{acond1}
\mbox{ (Fig. 7a)}:
\quad P&=&k_1+k_2,\; P'=k_1'+k_2', \nonumber \\
        P''&=&k_1''+k_2,\;P''=P+q,
\end{eqnarray}
\begin{eqnarray} \label{acond2}
\mbox{ (Fig. 7b)}:
\quad P&=&k_1+k_2,\; P'=k_1'+k_2', \nonumber \\
        P''&=&k_1''+k_2',\;P'=P''+q.
\end{eqnarray}

To treat the expressions
$$k_\mu,\; k_\mu',\; k_\mu k_\nu',\; k_\mu' k_\nu',$$
we have to represent $k_\mu,\; k_\mu'$ in the form
\begin{eqnarray} \label{kmu}
k_\mu=a_1 P_\mu&+&a_2 q_\mu+a_3\delta_\mu,\nonumber\\
k'_\mu=a'_1 P_\mu&+&a'_2 q_\mu+a'_3\delta_\mu,\nonumber\\
q&=&P'-P,\nonumber\\
\mbox{ (Fig. 7a)}:\quad\delta&=&P''-P',\nonumber\\
\mbox{ (Fig. 7b)}:\quad\delta&=&P''-P,\nonumber\\
a_1=\frac{(k\delta)}{(P\delta)}&,& \;
a_2=\frac{(k\delta)(Pq)}{(P\delta)Q^2}-\frac{(kq)}{Q^2},\nonumber\\
a'_1=\frac{(k'\delta)}{(P\delta)}&,&\;
a'_2=\frac{(k'\delta)(Pq)}{(P\delta)Q^2}-\frac{(k'q)}{Q^2}.
\end{eqnarray}

After isolating invariant amplitudes for which we write down the
dispersion relations over $s, s'$ and $s''$, the four-vectors should
be put on mass shell in the Lorentz tensor structures using the
conditions $P'=P+q$ (Fig.\ \ref{fig:loops}a) or $P''=P$ (Fig.
\ref{fig:loops}b).

To relate different scalar products, which appear during calculations
of traces of Eq.\ (\ref{sp1}, \ref{sp2}), with light--cone variables
we use the following four--vectors which are common for both diagrams
of Fig.\ \ref{fig:loops}a and \ref{fig:loops}b:
$$
k=\left(x P_+,\frac{m^2+\vec k_\perp^2}{2xP_+},\vec k_\perp\right),
k'=\left(x'P_+,\frac{m^2+\vec k_\perp'^2}{2x'P_+},\vec k_\perp'\right)
,$$
\begin{eqnarray} \label{lccom}
P'=\left(P_+,\frac{s' +Q^2}{2P_+},\vec Q \right)&,&
   q=\left(0  ,\frac{s'-s+Q^2}{2P_+}  ,\vec Q \right),\nonumber\\
P =\left(P_+\right.&,&\left.\frac{s}{2P_+},0  \right).
\end{eqnarray}
and four--vectors which are different for diagrams of
Fig.\ \ref{fig:loops}a and \ref{fig:loops}b:
\begin{eqnarray} \label{lcdif}
\mbox{ (7a)}:&& P''=\left(P_+,\frac{s''+Q^2}{2P_+},\vec Q \right),
 \delta=\left(0,\frac{s''-s'}{2P_+},0\right),\nonumber\\
\mbox{ (7b)}:&& P''=\left(P_+,\frac{s''}{2P_+},0  \right),
 \delta=\left(0,\frac{s''-s }{2P_+},0\right).
\end{eqnarray}

With Eq.\ (\ref{scond1}) (or Eq.\ (\ref{scond2})), the coefficients in
Eq.\ (\ref{kmu}) are equal to:
\begin{equation} \label{acoefs}
a_1=x, \quad a'_1=x',\quad a_2=\frac{(\vec k_\perp\vec Q)}{Q^2},\quad
a'_2=x'+\frac{(\vec k_\perp' \vec Q )}{Q^2}.
\end{equation}

Expansions of $k_\mu k_\nu'$ and $k_\mu' k_\nu'$ read as
\begin{eqnarray} \label{bexpres}
k_\mu k_\nu'&=&\sum\limits_{A,B=P,q,\delta}a'_{AB} A_\mu B_\nu
+b'g_{\mu\nu},\nonumber\\
k_\mu'k_\nu'&=&\sum\limits_{A,B=P,q,\delta}a''_{AB} A_\mu B_\nu
+b''g_{\mu\nu},
\end{eqnarray}
where $A_\mu=P_\mu,\; q_\mu,$ or $\delta_\mu$ and $B_\nu=P_\nu,\;
q_\nu,$ or $\delta_\nu$. For the calculation, we need coefficients
$b'$ and $b''$ only. For that one has to multiply both sides of Eq.\
(\ref{bexpres}) by $\Pi_{\mu\nu}$, which is orthogonal to each pair
$A_\mu B_\nu$, and $\Pi_{\mu\nu}g_{\mu\nu}=1$:
\begin{eqnarray} \label{PImunu}
\Pi_{\mu\nu}&=&g_{\mu\nu}
+\alpha_1 P_\mu P_\nu+\alpha_2 q_\mu q_\nu
+\alpha_3 \delta_\mu \delta_\nu \nonumber\\
&+&\beta_1( P_\mu q_\nu+P_\nu q_\mu)
+\beta_2(P_\mu\delta_\nu+P_\nu\delta_\mu)\nonumber\\
&+&\beta_3(q_\mu\delta_\nu+q_\nu\delta_\mu),
\end{eqnarray}
\begin{eqnarray}
\alpha_1=\frac{(q\delta)^2 - q^2\delta^2}D ,&&\quad
\beta_1=\frac{\delta^2(Pq) - (P\delta)(q\delta)}D,\nonumber\\
\alpha_2=\frac{(P\delta)^2 - P^2\delta^2}D ,&&\quad
\beta_2=\frac{q^2(P\delta) - (Pq)(q\delta)}D,\nonumber\\
\alpha_3=\frac{(Pq)^2 - P^2 q^2}D ,&&\quad
\beta_3=\frac{P^2(q\delta) - (Pq)(P\delta)}D,\nonumber\\
D=P^2q^2\delta^2+2(Pq)&&(P\delta)(q\delta)\nonumber\\
-P^2(q\delta)&&^2-q^2(P\delta)^2-\delta^2(Pq)^2.
\end{eqnarray}
Then,
\begin{eqnarray} \label{bcoefs}
b'&=&\Pi_{\mu\nu}k_\mu k'_\nu=\frac{(\vec k_\perp\vec Q)
(\vec k_\perp'\vec Q)}{Q^2}-(\vec k_\perp\vec k_\perp'),
\nonumber\\
b''&=&\Pi_{\mu\nu}k'_\mu k'_\nu=\frac{(\vec k_\perp'\vec Q)^2}{Q^2}
-\vec k_\perp'^2 .
\end{eqnarray}

Now we can write down final expressions for traces of Eqs.\
(\ref{sp1}, \ref{sp2}):
\begin{eqnarray} \label{sp1f}
Sp_{\mu\nu}^{(1)}=4m\varepsilon_{\mu\mu\alpha\beta}q^\alpha P^\beta
&&\left(\right.s'+4b''-12b'\nonumber\\
&&\left.+(s''-s-q^2)(a_1'-a_2')\right),
\end{eqnarray}
\begin{eqnarray} \label{sp2f}
Sp_{\mu\nu}^{(1)}=8m\varepsilon_{\mu\mu\alpha\beta}q^\alpha P^\beta
&&\left(\right.s-(a_1'-a_2')(k_1'' P)-2b''
\nonumber\\ &&\left.-a_2'(k_1'k_2')-m^2(1+a_2')\right),
\end{eqnarray}
where
\begin{eqnarray}
(k_1'' P)&=&\frac 12 (x's''+(1-x')s),\nonumber\\
(k_1'k_2')&=&\frac 12 s'-m^2.
\end{eqnarray}

\newpage
\begin{table}
\caption{Step--function approximation for  soft wave functions
($\vec k^2$ is given for the middle point of each step of Fig. 3).
All quantities are in $GeV$.}
\label{table:wf}
\begin{tabular}{|ccccc|}
$\vec k^2$ & $\psi_{n\bar n}$ & $\psi_{s\bar s}$ &
$\psi_{\gamma\to n\bar n}$ & $\psi_{\gamma\to s\bar s}$ \\
\hline
\hline
 0.033 & 87.737 & 42.316 & 14.488 &  7.949\\
 0.096 & 48.591 & 27.154 &  7.700 &  4.864\\
 0.159 & 28.170 & 17.194 &  5.006 &  3.445\\
 0.221 & 15.385 &  9.972 &  2.881 &  2.102\\
 0.284 &  9.005 &  6.096 &  2.438 &  1.855\\
 0.346 &  4.917 &  3.440 &  1.864 &  1.465\\
 0.440 &  2.216 &  1.609 &  1.369 &  1.116\\
 0.565 &  1.610 &  1.210 &  0.941 &  0.794\\
 0.690 &  1.480 &  1.140 &  0.608 &  0.526\\
 0.815 &  1.592 &  1.249 &  0.575 &  0.506\\
 0.940 &  1.905 &  1.517 &  0.602 &  0.538\\
 1.065 &  1.974 &  1.589 &  0.548 &  0.495\\
 1.190 &  2.227 &  1.810 &  0.560 &  0.510\\
 1.315 &  2.477 &  2.029 &  0.531 &  0.488\\
 1.440 &  2.307 &  1.902 &  0.516 &  0.477\\
 1.565 &  2.489 &  2.063 &  0.479 &  0.445\\
 1.690 &  2.029 &  1.690 &  0.473 &  0.442\\
 1.815 &  1.687 &  1.412 &  0.444 &  0.417\\
 1.940 &  1.525 &  1.281 &  0.446 &  0.420\\
 2.065 &  1.163 &  0.981 &  0.453 &  0.428\\
\end{tabular}
\end{table}

\end{document}